\documentclass[12pt]{article}
\usepackage{graphicx}
\begin{document}
\newcommand{\beq}{\begin{equation}}
\newcommand{\eeq}{\end{equation}}
\renewcommand{\theequation}{\thesection.\arabic{equation}}
\hoffset=-1.2truecm
\title{Comparison of exact and approximate cross-sections in 
relativistic
Coulomb excitation}
\author{B.F. Bayman\\
{\it School of Physics and Astronomy, University of Minnesota,}\\
{\it 116 Church Street S.E., Minneapolis, MN 55455, U.S.A.}\\
and\\
F.Zardi\\
{\it Istituto Nazionale di Fisica Nucleare and Dipartimento di Fisica,} 
\\
{\it Via Marzolo,8 I-35131 Padova,Italy.}}
\maketitle
\begin{abstract}

We present a new method of obtaining time-dependent matrix elements of
the electromagnetic pulse produced by a highly-relativistic projectile.
These matrix elements are used in a coupled-channel calculation to
predict the cross-sections for population of 1- and 2-phonon states of
the giant dipole resonance. Comparisons are made with the predictions 
of
the long-wavelength and Born approximations.
\end{abstract}

\section{Introduction}

The subject of relativistic Coulomb excitation has received extensive
study in the two decades since its theoretical foundations were 
established
in the classic work of Alder and Winther \cite{WA}. 
In the semiclassical approach to relativistic nuclear Coulomb 
excitation, the
relative motion of the projectile and target is treated classically. 
Indeed, it
is usually assumed to be straight-line motion at constant speed $v$. 
The
evolution of the internal degrees of freedom of each nucleus, under the 
influence
of the classical electromagnetic field produced by the other nucleus, 
is then
followed using quantum mechanics.

There exists a wide literature devoted to the use of relativistic 
Coulomb excitation
for the study of several aspects of nuclear structure 
that cannot be
explored through the nuclear interaction [2-8].
In recent years, 
this subject has been also shown to be relevant to the practical 
question 
of the stability of
beams of relativistic heavy ions, since some of the processes which 
lead to 
loss of beam ions are initiated by Coulomb excitation [2,9-11].

The situations that are easiest to interpret are those that can be 
described in terms of the Coulomb excitation of a single level. If this 
level is unbound, the Coulomb excitation will be followed by particle  
emission. Methods have been developed 
to extract from the particle emission cross-section the part that is 
associated with the Coulomb excitation of the emitting state, as 
opposed to excitation via nuclear forces \cite{HL76,ABS}. 
The Fermi-Williams-Weizs\"acker (FWW) 
method of virtual quanta \cite{JAC} is frequently used to analyze 
processes in 
which Coulomb excitation is followed by particle emission (see, e.g.,
refs.[5,12,15,16]). Here one simulates the electromagnetic pulse 
by an equivalent flux of virtual photons, and uses experimentally 
determined photo-nuclear cross-sections to describe the effect of these 
virtual photons on the target. However, most of the FWW applications to 
relativistic Coulomb excitation at energies of spectoscopic interest 
have not used \textit{experimentally} determined photo-nuclear cross-
sections, but have used \textit{calculated} electromagnetic matrix 
elements. Thus, in effect, these calculations really amount to the use 
of the Born approximation as developed in \cite{WA}, although they are 
expressed in the language of the method of virtual photons. 

Analyses of experimental data based on the Born approximation 
imply that the final state is weakly coupled to all states 
other than to the initial state. 
The Born approximation is just the first term of the Born (or Fredholm) 
series, which constitutes a formal solution of the time-dependent 
Schr\"odinger equation. In dealing with multi-channel processes, some 
workers \cite{GuWei} have attempted to include a few higher terms of 
this series. Other approaches, based on coupled-channel 
methods,\cite{BCH,LCC}, have attempted a numerical solution of the 
Schr\"odinger equation which, in principle, sums all terms of the Born 
series. The harmonic vibrator method allows for complete coupling of 
all the states 
of a vibrational band, if some simplifying assumptions are made 
concerning the ratios of interaction matrix elements.

All these calculations require matrix elements of the electromagnetic 
interaction between the nuclei, defined with respect to the eigenstates 
of internal motion. If the relative velocity of the nuclei is 
relativistic, the interaction is strongly retarded, which introduces 
difficulties into the evaluation of its matrix elements. The long-
wavelength approximation (LWLA) has proven to be a very useful device 
for overcoming these difficulties \cite{BCH,LCC}.

We see then that two approximations have been extensively used in this 
field: the Born approximation 
to solve the Schr\"odinger equation and the long-wavelength 
approximation 
to evaluate matrix elements. These approximations have been found to be 
effective below bombarding 
energies of 2 GeV per nucleon, which has been adequate for most 
nuclear structure investigations done so far. At higher energies, which 
could be of interest in future studies, and for energies presently 
considered in colliders, these approximations may not be adequate.

Much of the recent effort in the theoretical study of relativistic 
Coulomb excitation is devoted to the nuclear structure aspects of the 
problem. For example, in the study of giant resonances, one must take 
account of the spreading of the resonance amongst the background 
states (see, e.g., 
refs.\cite{BCH,BerPon99,PonMil,Pon687,Car99,Car99b}), 
and the 
anharmonicity of the oscillation \cite{ChFr,LCC,PfCar,Huss,deP}. 
Of course, these 
studies also must incorporate the interaction of the nuclear motion 
with the electromagnetic pulse, and this is generally done by means of 
the Born approximataion and/or the the LWLA. For this reason, we 
believe it is important to understand the limitations of these 
approximations, even in Coulomb excitation studies whose primary 
emphasis is on nuclear structure.  

The purpose of this article is to assess the ranges of validity of the 
LWLA and the Born approximation. We accomplish this by comparison with 
a general numerical solution of the time-dependent Schr\"odinger 
equation, expressed as a set of coupled differential equations. The 
energy range considered here extends from energies currently used for 
nuclear structure studies (see, \textit{e.g.} refs. [5,6]), up to about 
10 GeV per nucleon. The methods employed are quite general, and could 
be used at still higher energy. 

In Section 2, a general survey of methods and approximations is given. 
In Section 3, we present our method for numerical evaluation of the
Fourier transform of interaction matrix elements. In Sections 4 and 5, 
we
illustrate our procedure by investigating the relativistic Coulomb 
excitation
of giant dipole resonance (GDR) phonons. We present numerical results 
for
the particular case of  $^{208}$Pb projectiles exciting GDR 
states of $^{40}$Ca.
These results are compared with those produced by approximate 
approaches. We summarize in Section 6 the main results we obtained.

\section{Survey of methods used in RCE}
\setcounter{section}{2}
\setcounter{equation}{0}

In the semiclassical theory of relativistic Coulomb excitation, the
electromagnetic field between the projectile and target is treated 
classically, but their internal degrees of freedom are treated 
according to 
the principles of quantum mechanics.
The time dependent Schroedinger equation is written 
\beq
i \hbar \frac{\partial \psi(t)}{\partial t}~=~[H_{0}~+~V(t)]\psi(t)
\label{ii.1}
\eeq
with $H_0$ referring only to the target\footnote{For simplicity, we
restrict our discussion to the situation in which only the target is
excited, but the argument is easily generalized to allow for projectile
excitation as well.} degrees of freedom, and $V(t)$ 
the interaction between the target and the electromagnetic field of 
the projectile:
\beq
V(t)~=~\int~[\varphi_{_C}^{\rm ret}({\bf r'},t)\rho({\bf r'})
~-~\frac{1}{c}{\bf A}_{_C}^{\rm ret}({\bf r'},t)\cdot{\bf j}({\bf r'}
)]d^3r'~~.
\label{ii.2}
\eeq
The scalar and the vector potentials associated with the projectile
electromagnetic field are $\varphi_C^{\rm ret}({\bf r'},t),~
{\bf A}_C^{\rm ret}({\bf r'},t)$, and $\rho({\bf r'}),~{\bf j}({\bf r'
})$are the target transition charge and current densities, 
respectively.

The projectile is assumed to have a spherically symmetric charge 
distribution, with total charge $Z_Pe$. Because of its large momentum, 
it can be 
assumed to follow a straight-line trajectory at constant speed $v$.
Thus its center is located at time $t$ by 
\beq
{\bf r}~=~b{\bf{\hat y}}~+~vt{\bf{\hat z}}~~.
\label{ii.3}
\eeq
The constant $b$ is the impact parameter. Excitation probabilities are
calculated as functions of $b$. The scalar 
and vector potentials of the projectile field, at the target point
${\bf r'}$, are given by \cite{JAC}
\begin{eqnarray}
\varphi_{_C}^{\rm ret}({\bf r'},t)&=&{Z_{_P}e\gamma \over 
\sqrt{x'^{2}+(y'-b)^{2}+\gamma^{2}(z'-vt)^{2}}}\nonumber \\
{\bf A}_{_C}^{\rm ret}({\bf r'},t)&=&\frac{v}{c}~
\varphi_{_C}^{\rm ret}({\bf r'},t)
{\bf{\hat z}}
\label{ii.4}   
\end{eqnarray}
Because of the factor $\gamma^2$ in the denominator of Eq.(2.4), these
potentials are not spherically symmetric about the projectile center.

The Schroedinger equation (2.1) is conveniently expressed in terms 
of an expansion in eigenstates $\phi_\alpha$ of the nuclear Hamiltonian
$H_0$,
\beq
\psi(t)~=~\sum_\alpha~a_\alpha(t)e ^{\frac{i}{\hbar}
\epsilon_\alpha t}\phi_\alpha
~~,
\eeq
with
\beq
H_0\phi_\alpha~=~\epsilon_\alpha\phi_\alpha~~,~~~~~~~~
<\phi_\alpha|\phi_\beta>~=~\delta_{\alpha\beta}~~;
\eeq
\beq
V_{\alpha\beta}(t)~=~<\phi_\alpha|\varphi_C^{\rm ret}(t)\rho
~-~\frac{1}{c}{\bf A}_C^{\rm ret}(t)\cdot{\bf j}|\phi_\beta>~~.
\eeq
If (2.5) is substituted into (2.1), the result is a set of coupled 
ordinary differential equations for the amplitudes $a_\alpha(t)$:
\beq
i\hbar~\frac{da_\beta(t)}{dt}~=~\sum_{\alpha}~
e^{\frac{i}{\hbar}
(\epsilon_\beta-\epsilon_\alpha)t}~V_{\beta\alpha}(t)a_\alpha(t)~~.
\eeq
These equations must be solved subject to initial conditions
$$
a_\alpha(-\infty)~=~\delta_{\alpha,0}~~.
$$

The probability that the target will be in state $\phi_\beta$ after 
the collision is given by $|a_\beta(+\infty)|^2$, and the cross-section
for the population of $\phi_\beta$ is given by
\beq
\sigma_\beta~=~\int_{b_{\rm min}}^\infty~2\pi|a_\beta(+\infty)|^2 
bdb~~.
\eeq
Here $b_{\rm min}$ is usually taken to be somewhat larger than the sum
of the target and projectile radii\cite{EML}.

The first step in the solution of Eq.(2.8) is the computation of
the interaction matrix elements $V_{\beta\alpha}(t)$. This is 
a formidable task, since the retardation of the interaction introduces 
a 
directional
asymmetry, which means that the interaction is not invariant under
rotations of the internal coordinates alone. 
Fortunately, Alder and Winther have found a convenient multipole
expansion of the {\it Fourier transform} of the $V_{\beta\alpha}(t)$,
\beq
V_{\beta\alpha}(\omega)~\equiv~\int_{-\infty}^\infty\frac{dt}{\hbar}
e^{i\omega t}V_{\beta\alpha}(t)~~.
\eeq
The $V_{\beta\alpha}(\omega)$ defined in Eq.(2.10) can be used directly
in a coupled integral equation formulation of the Schroedinger 
equation, 
as was done in ref.\cite{BZ}.
However this 
approach is difficult to implement when many states $\phi_\alpha$ have
to be included in the calculation. Another alternative is to proceed by
inverting the Fourier transform (2.10) 
\beq
V_{\beta\alpha}(t)~=~\frac{\hbar}{2\pi}\int_{-\infty}^\infty d\omega
e^{-i\omega t}V_{\beta\alpha}(\omega)
\eeq
to convert the $V_{\beta\alpha}(\omega)$ obtained from the Alder-
Winther
expansion into $V_{\beta\alpha}(t)$ which can be used in the time-
dependent
formulation (2.8) of the Schroedinger equation.

There is no known closed expression for the Fourier transform (2.11).
In ref.\cite{LCC} Lanza {\it et al} replaced the exact Alder-Winther
expression for $V_{\beta \alpha}(\omega)$ by its long-wavelength
approximation (LWLA). Here one assumes that the important Fourier
components of the electromagnetic pulse correspond to wavelengths that
are large compared to target dimensions. In this case, the
small-argument limit of the spherical Bessel function,
\beq
j_{\lambda}(\frac{\omega}{c}r')\sim\frac{(\frac{\omega}{c}r')^{\lambda}
}
{(2 \lambda+1)!!}, 
\eeq
can be used.
With this approximation, Lanza {\it et al} \cite{LCC} 
were able to obtain explicit approximate expressions, in terms of 
hypergeometric functions, for the $V_{\beta \alpha}(t)$,
which they used in their analyses of multiphonon Coulomb excitation.
They tested the validity of the LWLA by
comparing a few $V_{\beta \alpha}(t)$ calculated using the LWLA with
numerical evaluations of the Fourier transform (2.11) of the exact
$V_{\beta \alpha}(\omega)$. They concluded that the LWLA was adequate
for their analysis of the $^{208}$Pb+$^{208}$Pb collision at E$_{{\rm
lab}}$=641 MeV. 

Bertulani {\it et al} \cite{BCH} approached the problem of finding
approximate $V_{\beta \alpha}(t)$ by expressing the potential (2.4) in 
a 
Taylor expansion around $x'=y'=z'=0.$ This formally exact expansion was
truncated in a manner that caused the remaining terms to be precisely
equal to those given by the LWLA expressions in ref.\cite{LCC}. 
However, 
the advantage of the truncated Taylor expansion is
that it yields simple expressions for the $V_{\beta \alpha}(t)$ which,
although they are equal to the hypergeometric functions used in 
ref.\cite{LCC}, are much more transparent. The general proof of the
equivalence of these methods can be found in ref.\cite{BZ03}.

The approach to be followed in this manuscript is the use of a quick 
and
accurate method for the numerical evaluation of the Fourier tranform
(2.11), using exact $V_{\beta \alpha}(\omega)$. The
$V_{\beta \alpha}(t)$ calculated in this way will be used in the
numerical solution of the coupled time-dependent equations (2.8). We
will thus be able to obtain essentially exact Coulomb-excitation
cross-sections at any bombarding energy. These exact cross-sections
can be used to
explore the limits of validity of the LWLA and other approximations.

\section{Numerical Evaluation of the Fourier transform}
\setcounter{section}{3}
\setcounter{equation}{0}

The difficulty of evaluating (2.11) as a numerical integral is the 
rapid 
oscillation of the integrand for high values of $t$. Our approach to
this problem is a generalization of the idea behind the use of 
Simpson's
rule in the evaluation of $\int_{x_i}^{x_f}f(x)dx$. We first divide the
$\omega$-integration range of (2.11) into an even number of steps of
length $h$. We assume that these steps are small enough so that, over
every adjacent pair of steps, $V_{\beta \alpha}(\omega)$ can be
approximated by a quadratic function of $\omega$. Thus for the interval
$\omega_{1}-h\le\omega\le\omega_{1}+h$, we make the approximation
$$
V_{\beta \alpha}(\omega)\sim(\frac{\omega-\omega_{1}}{h})^2~u~+
~(\frac{\omega-\omega_{1}}{h})~v~+w
$$
We can make this approximation exact at
$\omega=\omega_{1},~\omega_{1}\pm h$ by defining $u,v,$ and $w$ to be
\begin{eqnarray*}
u&=&\frac{V_{\beta\alpha}(\omega_1-
h)~+~V_{\beta\alpha}(\omega_1+h)}{2}~-~
V_{\beta\alpha}(\omega_1)\nonumber\\
v&=&\frac{V_{\beta\alpha}(\omega_1+h)~-~V_{\beta\alpha}(\omega_1-h)}{2}
\nonumber\\
w&=&V_{\beta\alpha}(\omega_1)~~~.
\end{eqnarray*}
Note that since the quadratic approximation is applied to 
$V_{\beta\alpha}(\omega)$, its validity is independent of $t$. Then the 
$\omega_{1}-h\le\omega\le\omega_{1}+h$ part of the integration (2.11)
can be approximated by
\beq
\int_{\omega_1-h}^{\omega_1+h} d\omega e^{-i\omega
t}V_{\beta\alpha}(\omega)\sim\int_{\omega_1-h}^{\omega_1+h} d\omega 
e^{-i\omega
t}~[~(\frac{\omega-\omega_{1}}{h})^2~u~+~(\frac{\omega-
\omega_{1}}{h})~v~+w~]
\eeq
Now an exact integration of the right-hand side of (3.1) yields
\begin{equation}
\int_{\omega_1-h}^{\omega_1+h}~d\omega e^{-i\omega 
t}V_{\beta\alpha}(\omega)
\simeq
\end{equation}
$$
\frac{2}{h^2t^3}\Bigl\{~ht\cos(ht)\Bigl(2u\cos(\omega_1t)+vht
\sin(\omega_1t)~\Bigr)
$$
$$
+\sin(ht)\Bigl(~(-2u+h^2t^2(u+w))\cos(\omega_1t)-
vht\sin(\omega_1t)~\Bigr)\Bigr\}
$$
$$
-\frac{2i}{h^2t^3}\Bigl\{~ht\cos(ht)\Bigl(2u\sin(\omega_1t)-vht
\cos(\omega_1t)~\Bigr)
$$
$$
+\sin(ht)\Bigl(~(-2u+h^2t^2(u+w))\sin(\omega_1t)+
vht\cos(\omega_1t)~\Bigr)\Bigr\}
$$
This approximation is used for every adjacent pair of steps in the
$\omega$ integration. 

Equation (3.2) cannot be used at $t$=0, since it takes the form 0/0 
there.
By expanding the quantities between \{~\} about $t$=0, we
can show that  
$$
\lim_{t\to0}\int_{\omega_1-h}^{\omega_1+h}~d\omega e^{-i\omega t}
V_{\beta\alpha}(\omega)
$$
\begin{eqnarray}
&&\simeq\frac{2}{3}h(u+3w)-\frac{t^2}{15}\Bigl(h^2(3u+5w)+10hv\omega_1+
5(u+3v)\omega_1^2\Bigr)\nonumber\\
&&-\frac{2i}{3}ht\Bigl(~hv+(u+3w)\omega_1\Bigr)
\end{eqnarray}

The only circumstance where the quadratic representation of
$V_{\beta\alpha}(\omega)$ is inadequate occurs when 
$V_{\beta\alpha}(\omega)$
is singular at $\omega$=0. In these cases 
$V_{\beta\alpha}(\omega)$ has the form
\beq
V_{\beta\alpha}(\omega)~=~f(\omega)~K_0(\frac{|\omega|b}{\gamma v})~~,
\eeq
with $f(0)\neq0$. Since $K_0(\frac{|\omega|b}{\gamma v})$ diverges 
logarithmically as $\omega\to0$, numerical integration of (3.4) 
requires
special precautions. In this case, we work with the identity
\beq
V_{\beta\alpha}(\omega)~=~\Bigl(f(\omega)~-
~f(0)\Bigr)~K_0(\frac{|\omega|b}
{\gamma v})~+~f(0)~K_0(\frac{|\omega|b}{\gamma v})~~.
\eeq
The first term in (3.5) is regular at $\omega=0$, and its Fourier 
transform
can be evaluated without difficulty using Eqs.(3.2) and (3.3). The 
Fourier
transform of the second term in (3.5) can be evaluated exactly
\beq
\frac{\hbar}{2\pi}\int_{-\infty}^\infty~d\omega~e^{-i\omega t}f(0)
K_0(\frac{|\omega|b}{\gamma v})~=~\frac{f(0)\hbar v}
{2\sqrt{(\frac{b}{\gamma})^2+(vt)^2}}~~.
\eeq
This term dominates the interaction matrix element $V_{\beta\alpha}(t)$
at large values of $|t|$.

\section{Brink's model for the Giant Dipole Resonance}
\setcounter{section}{4}
\setcounter{equation}{0}
\subsection{The GDR phonon states}

The giant dipole resonance can be regarded as a collective oscillation 
of the
protons in a nucleus relative to the neutrons \cite{GTel}. The
one-phonon GDR state also has a simple interpretation in terms of an
isovector linear combination of one-particle, one-hole excitations of 
the
ground state \cite{Wil}. It was shown by Brink \cite{Bri}
that states of the GDR have a simple interpretaion within the harmonic 
oscillator shell model, in terms of harmonic oscillations of the vector 
${\bf R}_{{\rm pn}}$ connecting the centers of unexcited proton and 
neutron spheres. We use this representation of the GDR to
provide the transition charge and current densities needed for the
calculation of the $V_{\beta \alpha}(\omega)$ (see App. A).

An individual GDR state is described by the three quantum numbers 
$N,L,M,$ in
which $N(=0,1,2\cdots)$ is the principal quantum number. The number of 
GDR
phonons associated with this state is $2N+L$. Thus the $\beta,\alpha$ 
in  
$V_{\beta \alpha}(\omega)$ stand for $N_{\beta},L_{\beta},M_{\beta}$ 
and 
$N_{\alpha},L_{\alpha},M_{\alpha}$, respectively. These states are 
denoted $\Psi^{NL}_M$.

\subsection{Structure of the matrix elements}

The algebraic details of the construction of the GDR states, and the
derivations of the formulae for the associated $V_{\beta 
\alpha}(\omega)$, are
given in the Appendices. In order to facilitate our discussion of the 
long-wavelength approximation, and of the features that lead to 
convergence of the $\omega$ integral, we summarize here the general 
structure of the result. 

$V_{\beta \alpha}(\omega)$ can be expressed in terms of the Fourier 
transforms of the scalar and vector potentials as follows:  
\beq
V_{\beta \alpha}(\omega)~=~<\Psi^{N_\beta L_\beta}_{M_\beta}|V(\omega)|
\Psi^{N_\alpha L_\alpha}_{M_\alpha}>
~=~\int d^3r'~[~\rho_{\beta \alpha}({\bf r'})\varphi_{_C}^{\rm 
ret}({\bf
r'},\omega)
~-~\frac{1}{c}{\bf j}_{\beta \alpha}({\bf r'}) \cdot 
{\bf A}_{_C}^{\rm ret}({\bf r'},\omega)] ~~,
\eeq
where
$$
\varphi_{_C}^{\rm ret}({\bf r'},\omega)=\int_{-
\infty}^\infty\frac{dt}{\hbar}
e^{i\omega t}~\varphi_{_C}^{\rm ret}({\bf r}',t)
$$
\beq
{\bf A}_{_C}^{\rm ret}({\bf r'},\omega)=\frac{v}{c}
\varphi_{_C}^{\rm ret}({\bf r'},\omega){\hat{\bf z}}
\eeq

The multipole expansion of $\varphi_{_C}^{\rm ret}({\bf r}',\omega)$, 
given first by Alder and Winther \cite{WA}, can be 
written in the form
\beq
\varphi_{_C}^{\rm ret}({\bf r}',\omega)~=~\sum_{\lambda,\mu}~
K_\mu(\frac{|\omega|b}{\gamma v})~
C_{\lambda,\mu}
(\omega)~j_{\lambda}(\frac{|\omega|}{c}r')~
Y^{\lambda}_{\mu}({\bf {\hat r'}})
\eeq
with
$$
C_{\lambda,\mu}(\omega)~\equiv~e^{-
i\mu\frac{\pi}{2}}~\frac{2Z_Pe}{\hbar v}~
{\cal G}_{\lambda,\mu}
$$
where
$$
{\cal G}_{\lambda,\mu}~=~\frac{i^{\lambda+\mu}}{(2 
\gamma)^{\mu}}~(\frac{\omega}{|\omega|})^{\lambda-
\mu}~(\frac{c}{v})^{\lambda}
~\times~\sqrt{4 \pi~ (2\lambda+1)~(\lambda-
\mu)!~(\lambda+\mu)!}
~\sum_{n}\frac{1}{(2 \gamma)^{2n}(n+\mu)!n!(\lambda-\mu-2n)!}
$$

It is shown in Appendix B that the $\int\varphi\rho$ part of the matrix 
element can be expressed in the form:
\beq
\int d^3r'~\varphi_{_C}^{\rm ret}({\bf r'},\omega)
\rho_{\beta \alpha}({\bf r'})
~=~\sum_{\lambda
\mu}C_{\lambda,\mu}(\omega)K_\mu(\frac{|\omega|b}{\gamma v})~\int 
d^{3}r'~
\rho_{\beta \alpha}({\bf
r'})j_{\lambda}(\frac{|\omega|}{c}r')Y_{\mu}^{\lambda}({\hat r'})
\eeq
\beq
\simeq e^{\frac{(\omega/c)^2}{8
Z_T\nu}}\sum_{n',\ell'}2(2\ell'+1)\int_0^\infty~j_0(\frac{\omega}{c}r')
u_{n',\ell'}^2(r')r'^2~dr'~
\sum_{n\ell NL}~A_{n\ell NL}
~K_\mu(\frac{|\omega|b}{\gamma v}){\tilde u}_{NL}(\frac{\omega/c}
2\sqrt{2}),
\eeq
where $A_{n\ell NL}$ are suitable coefficients.
The functions $u_{n\ell}(r)$ are 
the harmonic oscillator radial functions associated with the individual
nucleons moving in the shell-model potential,
\beq
u_{n \ell}(r)=\sqrt{\frac{2^{\ell-n+2}(2 
\ell+2n+1)!!~\nu^{3/2}}{\sqrt{\pi}n!
}}~(\sqrt{\nu}r)^{\ell}~e^{- \frac{\nu r^2}{2}}
\times\sum_{\kappa}~\frac{(-2)^\kappa~n!}{\kappa
!(n-\kappa)!(2 \ell+2 \kappa +1)!!}~(\nu r^2)^{\kappa},
\eeq
where $\nu=m\omega_{{\rm sm}}/\hbar$, with $\omega_{{\rm sm}}$ 
representing the frequency associated with the shell-model harmonic 
oscillator 
potential. The tilde in the corresponding Fourier transformed states
${\tilde u}$ signifies that the size parameter is 
taken to be $2/(Z_T \nu)$. The matrix elements of the $-\frac{1}{c}
\int{\bf j}_{\beta \alpha}\cdot {\bf A}_{_C}^{\rm ret}$ part of the 
electromagnetic interaction are easily 
represented in terms the above formulae (see Appendix C).

The advantage of this approach is that it yields explicit functions for 
the
matrix elements $V_{\beta \alpha}(\omega)$, with only the single 
parameter
$\nu$ (which is related to the nuclear radius). Of course, the 
representation
of GDR states in terms of eigenstates of the independent particle shell 
model
with oscillator radial functions is highly schematic. Some important 
features
of the real GDR, such as the spreading of the GDR among background 
states, are
missing from this model. However, the states we use give a realistic 
picture of
the collective oscillation of the nuclear protons and neutrons relative 
to each
other, and we believe they provide sufficiently accurate $V_{\beta
\alpha}(\omega)$ to enable us to test the dynamics of the Coulomb 
excitation
process. 

\subsection{The Long-Wavelength Approximation (LWLA)}

\subsubsection{Expression of the LWLA in Brink's model of the GDR}

We can get the result of using the long wavelength approximation (2.12) 
in (4.4) by keeping
only the lowest powers of $\frac{|\omega|}{c}$ in (4.5), apart from the
$\omega$ dependence of $ K_{|M_\beta-M_\alpha|}(\frac{|\omega|b}{\gamma 
v})$. 
This implies the following replacements in (4.5):
$$
e^{\frac{(|\omega|/c)^2}{8Z\nu}}~\longrightarrow~1
$$
$$
\int_0^\infty j_0(\frac{|\omega|}{c}r')u_{n,
\ell}^{2}(r')r'^2dr'\longrightarrow~\int_0^\infty u_{n,
\ell}^{2}(r')r'^2dr'~=~1
$$
$$
{\tilde u}_{NL}(\frac{|\omega|/c}{2
\sqrt{2}})~\longrightarrow~\sqrt{\frac{2^{L-
N+2}(2L+2N+1)!!}{(Z\nu/2)^{3/2}\sqrt{\pi}
[(2L+1)!!]^{2}}}~\left(\frac{|\omega|/c}{2\sqrt{Z\nu}}\right)^{L}
$$
If these replacements are made, our calculated results are equal to
those obtained with the LWLA. In Section 5 we will test the 
adequacy of the long-wavelength
approximation, for varying bombarding energies and impact parameters.

\subsubsection{General conditions for the validity of the LWLA}

If the integral representation (2.11) of $V_{\beta
\alpha}(t)$ is to converge, it is necessary that $V_{\beta
\alpha}(\omega)$ should decrease sufficiently rapidly as
$|\omega|\rightarrow \infty$. We use the term {\it damping} to refer to
the phenomena responsible for this decrease. It is evident from (4.4) 
that the decrease of $V_{\beta
\alpha}(\omega)$ with increasing $|\omega|$ has two causes:

1) {\it Kinematic damping}. The factor $K_\mu(\frac{|\omega|b}{\gamma 
v})$
has a large-$|\omega|$ limit of
$$
K_\mu(\frac{|\omega|b}{\gamma v})~~\stackrel{|\omega|>>\frac{\gamma
v}{b}}{\longrightarrow}~~\sqrt{\frac{\pi}{2(\frac{|\omega|b}{\gamma
v})}}~e^{-\frac{|\omega|b}{\gamma v}}
$$
This decrease occurs because the time-width of the electromagnetic
pulse at the target is of the order of $(\frac{b/\gamma}{v})$, which
contains frequency components up to, but not greatly exceeding,
$\omega_{{\rm max}}\sim \frac{\gamma v}{b}$. It can also be said that
the impulse is too {\it adiabatic} to contain frequency components with
$\omega > \omega_{{\rm max}}$.

2) {\it Dynamic damping}. If $\frac{|\omega|}{c}R>>1$, the factor $j_
{\lambda}(\frac{|\omega|}{c}r')$ in (4.4) will go through many
oscillations as $r'$ is integrated from 0 to $R$. These oscillations 
will lead
to cancellations between different parts of the $r'$ integration range, 
which
will lead to a decrease of  $V_{\beta \alpha}(\omega)$ when 
$|\omega|>c/R$. This
phenomena can also be called {\it retardation} damping, because it is 
associated
with the finite speed, $c$, of propagation of electromagnetic signals. 
Because
of this finite speed, different parts of the target nucleus, at one 
time
$t$, are influenced by the projectile at different points along its
orbit. This implies that the $r'$ integration needed for the evaluation
of $V_{\beta \alpha}(\omega)$ effectively produces a time average over
the projectile history, which smooths out sharper features of this
history. Thus $V_{\beta \alpha}(t)$ will not vary as rapidly with $t$ 
as
it would if there had been no retardation. Equivalently, the presence 
of
high-$|\omega|$ components in $V_{\beta \alpha}(\omega)$ is diminished.

It is worth remarking that the criterion for dynamic damping, 
$|\omega|>c/R$, is opposite to the validity condition for the LWLA, 
$|\omega|<c/R$. Thus if we remain within the regime of validity of the
LWLA, we will not experience dynamic damping; in this regime, the only 
phenomenon that can
lead to the damping required for the convergence of the integral (2.11)
is kinematic damping. Hence we conclude that the LWLA can only be 
applied
if $|\omega|<c/R$ for the {\it entire} $|\omega|$-range from $\omega=0$ 
up to
$\omega_{{\rm max}}\sim \frac{\gamma v}{b}$, the frequency at which
kinematic damping becomes effective. This implies that 
$$
\omega_{{\rm max}}\sim \frac{\gamma v}{b}~<~\frac{c}{R}~,
$$
$$
\frac{b}{R}~>~\frac{\gamma v}{c}~=~\sqrt{\gamma^2-1}~,
$$
must be true if the LWLA is to be trusted.

\section{Numerical results in 
$^{40}$C\lowercase{a}--$^{208}$P\lowercase{b} collisions}
\setcounter{section}{5}
\setcounter{equation}{0}
\subsection{Comparison of excitation amplitudes}

The coupled differential equations (2.8) for the $a_\beta(t)$ were
numerically integrated using the fourth-order Runge-Kutta method. We 
have
included all states with $\le$ two phonons, {\it i.e.} the states with
$(NLM)=(000),(010),(011),(020),(021),(022),(100)$ ($M=1,2$ here
represents the reflection-symmetric combinations of the $M=\pm1,\pm2$ 
states.) A test of the numerical accuracy of the integration procedure
is the extent to which the total normalization of the state,
 $\sum_\beta|a_\beta(t)|^2=1$, is preserved. It was not difficult to
satisfy this condition to 8 decimal places, even in situations when 
there
was strong coupling between the different $a_\beta(t)$.

The only free parameter in our model is the size parameter,
$\nu=m\omega_{{\rm sm}}/\hbar$, that characterizes the shell-model 
potential. 
For simplicity, we follow the usual prescrition
$$
\hbar \omega_{{\rm sm}} \sim 40 A^{-\frac{1}{3}}~~{\rm MeV},
$$
which, in the case of $^{40}$Ca, leads to $\hbar \omega_{{\rm sm}}=$ 
11.7 MeV, $\nu=.28177$ fm$^{-2}$.

We illustrate the criterion of validity of the LWLA with the example of 
$^{208}$Pb projectiles
of kinetic energy 10A GeV, on a $^{40}$Ca target. Then
$\gamma=1+10/.938 \sim 11.66$, and the condition of validity of LWL is
$$
b~>~\sqrt{(11.66)^2-1}\times R~\sim~50~{\rm fm}~.
$$

Figures 1a and 1b show $V_{\beta\alpha}(\omega)$ and
$V_{\beta\alpha}(t)$ for $b=12$ fm, with $\alpha$ referring to the
ground state and $\beta$ referring to the one-phonon state with $M=1$. 
It
is evident that the exact expression for $V_{\beta \alpha}(\omega)$, 
which
includes dynamic as well as kinematic damping, has less high-$|\omega|$
content than does the LWLA expression. Correspondingly, $V_{\beta
\alpha}(t)$ calculated with the LWLA is more sharply peaked than is the
exact $V_{\beta \alpha}(t)$. Figures 2a and 2b show the same 
comparison,
but for $b=50$ fm, where the above argument suggests that the LWLA 
should
be adequate. While it is clear that the LWLA does a better job at 
$b=50$ fm than at $b=12$ fm, the approximation at $b=50$ fm is still 
only fair.

The calculation of the total excitation probability requires an
integration over $b$. Thus the inadequacy of the LWLA for
small $b$ will have serious consequences only if the small-$b$ part of
the integration range makes an important contribution to the total
$b$-integration. We will see in Section 5 that the bombarding energy
determines whether or not this occurs.

Some typical plots of $P_\beta(t)\equiv|a_\beta(t)|^2$ are shown in 
Figures 3 through 6.

Figure 3 shows the occupation probability of $\Psi^{01}_{1}$ for an
impact parameter of 200 fm. The horizontal line indicates the Born
approximation prediction. It is seen that, beginning near $t$=0 fm/c, 
the
occupation probability rises almost monotonically to the Born value, 
and
then remains there. The occupation probability of $\Psi^{01}_{0}$ shown
in Figure 4 has a very different behavior. The coupling between 
$\Psi^{00}_{0}$ and $\Psi^{01}_{0}$ has a long range, because the
$\lambda=\mu=0$ term in the ${\bf j \cdot A}$ part of the interaction 
can
connect these states. It is seen in Figure 4 that the occupation
probability rises to a maximum near $t=0$ fm/c, and then strongly
decreases. The asymptotic probability in this case is 0.6 $\times
$ 10$^{-5}$, compared to the
Born prediction of $\sim 10^{-8}$. The rise and fall of the
$\Psi^{01}_{0}$ occupation probability is associated with the fact that
$\Psi^{01}_{0}$ represents a longitudinal oscillation (with reference 
to
the direction of the projectile motion), and the coupling to the ground
state changes sign during the encounter. $\Psi^{01}_{1}$, on the other
hand, is a transverse oscillation, and the coupling to the ground state
has a constant sign.

The Born prediction for the excitation probablility of $M=0$ states is
always very small at high bombarding energy (see Section 5.3.1 below). 
 
The corresponding curves for a grazing collison ($b=12$ fm) are shown 
in
Figures 5 and 6. Again the $\Psi^{01}_{1}$ occupation probability rises
almost monotonically to a constant value, but in this case the constant
value is significantly less than the Born probability. The population 
of
$\Psi^{01}_{0}$ shows a rather complicated time structure. The
asymptotic occupation probability is again much greater than the Born
probability.

The small-amplitude oscillation of the occupation probability of
$\Psi^{01}_{0}$ shown in Figure 6 for $b=12$ fm and $t>100$ fm/c is a
consequence of the long range of the $\lambda=\mu=0$ coupling term.

\subsection{Evaluation of the cross section}

Figures 7 and 8 show the asymptotic probabilites for $\Psi^{01}_{0}$
and $\Psi^{01}_{1}$, plotted as functions of impact paramater $b$.

In
the $\Psi^{01}_{1}$ case, it is seen that the Born expression gives a
very good representation of the exact probability when $b
\stackrel{>}{\sim}100$ fm. In this situation, the most convenient way 
to 
evaluate the cross-section is to break the impact-parameter integral 
(2.9) 
into two terms:
$$
\sigma_\beta~=~\int_{b_{{\rm min}}}^{b'}2 \pi
|a_\beta(+\infty)|^2 bdb~+~\int_{b'}^{\infty}2 \pi
|a_\beta(+\infty)|^2 bdb,
$$
where $b'\stackrel{>}{\sim}100$ fm. The first integral is evaluated
numerically using $|a_\beta(+\infty)|^2$ calculated by integrating the
Schrodinger equation for a range of $b$ values between $b_{{\rm min}}$ 
and $b'$. The second integral can be calculated exactly,
because the $b$ dependence of the Born probability is given by
$(K_{\mu}(\frac{|\omega|b}{\gamma v}))^2$, and one can take advantage 
of the
exact formula
$$
\int_{\xi}^{\infty}(K_{\mu}(x))^{2}xdx
=\frac{\xi^{2}}{2}~[~
(K_{\mu+1}(\xi))^{2}-(K_{\mu}(\xi))^{2}-\frac{2 \mu}{\xi}
K_{\mu+1}(\xi)K_{\mu}(\xi)~]
$$
Thus the contribution to the cross-section from 
$b'\le b<\infty$ will be 
$$
\sigma=\pi q^{2} b'^{2}~\frac{\xi^{2}}{2}~[~
(K_{\mu+1}(\xi))^{2}-(K_{\mu}(\xi))^{2}-\frac{2 \mu}{\xi}
K_{\mu+1}(\xi)K_{\mu}(\xi)~]
$$
where
$$
\xi\equiv\frac{b'\omega_{{\rm on-shell}}}{\gamma v}
$$
and
$$
q^{2}\equiv\frac{{\rm Born~probability~at~b'}}{(K_{\mu}(\xi))^{2}}
$$

In a situation such as that shown in Figure 8, where the Born formula
does not give an adequate approximation to the exact result even at
large $b$, it is necessary to do a numerical integration of (2.9) over
the entire $b$ range.

\subsection{Comparison of cross sections}

We use the term ``exact" to refer to calculations in which the
interaction matrix elements are evaluated without approximation, and
full account is taken of coupling between 0-, 1-, and 2-phonon states.
The LWLA calculations also include full coupling between 0-, 1-, and 2-
phonon states, but the interaction matrix elements are evaluated
approximately (Equation 2.12, and Section 4.3). The Born approximation
uses the correct on-shell interaction matrix elements, but assumes that
the transition to any final state is accomplished in a direct, single-
step process. Of course, our ``exact" calculation is not really exact,
since it leaves out the effect of coupling with states of 3, 4,
$\ldots$ phonons. For this reason, we have chosen to restrict our
attention to bombarding energies of $\le 10A$ GeV, with the expectation
that below this energy a description involving only 0-, 1-, and
2-phonons will be adequate. This is probably a fairly good 
approximation for 
the excitation cross-sections of the 1-phonon states. It is difficult 
to
assess the effect of our $\ge 3-$phonon truncation on the calculation 
of
the 2-phonon cross-sections. This will be investigated in a future
publication.

\subsubsection{The Born approximation}

A striking disagreement between the predictions of the Born 
approximation and the exact calculation is displayed in Figure 9. At 
bombarding energy below about 2A GeV, the Born approximation and exact 
calculation yield nearly the same cross-section for population of the 
1-phonon $M=0$ state. However, at higher bombarding energy, the Born 
approximation cross-section becomes vanishingly small, whereas the 
exact calculation predicts appreciable cross-section. It must be 
recalled that the Born approximation involves only the \textit{on-
shell} Fourier component of the interaction, $V_{\beta 
\alpha}(\omega=\frac{E_\beta-E_\alpha}{\hbar})$. It was shown in 
Reference \cite{BZ} that $\Delta M=0$ on-shell interaction matrix 
elements vanish at high bombarding energy in proportion to 
$1/\gamma^2$, as a result of cancellation between the contributions of 
the scalar and vector potentials. This accounts for the strong decrease 
at high bombarding energy of the Born $\Delta M=0$ cross-section. 
However, off-shell interaction matrix elements, $V_{\beta 
\alpha}(\omega \neq \frac{E_\beta-E_\alpha}{\hbar}),$ do not vanish at 
high $\gamma$; rather they diverge in proportion to $\log \gamma$. In 
the exact calculation, off-shell interaction matrix elements contribute 
to the population of the one-phonon $M=0$ state, as do multistep 
processes. Thus the strong Born approximation selection rule against 
population of the 1-phonon $M=0$ state is not exhibited by the exact 
calculation.

On the other hand, Figure 10 shows that the Born approximation and 
exact calculations are in fairly good agreement for the population of 
the 1-phonon $|M|=1$ state over the entire energy range. This suggests 
that most of the population of this state occurs in a single-step 
process. The fact that the exact calculation predicts a slightly 
smaller cross-section can be interpreted as a result of the loss of 
flux from the $|M|=1$ state associated with the coupling to other 
available states.

Figure 11 shows the cross-section for the total population of the 1-
phonon level, including both the $M=0$ and the $|M|=1$ states. It is 
seen that as a result of the Born approximation underprediction for the 
$M=0$ state, and over-prediction for the $|M|=1$ state, there is fairly 
good agreement for the total 1-phonon cross-section over the entire 
energy range. Thus measurements of the total 1-phonon cross-section 
cannot distinguish between the predictions of the Born approximation 
and exact calculations, and our exact calculation has little to 
contribute to the many analyses of total 1-phonon cross-sections 
performed with the Born approximation. However, if the separate $M=0$ 
and $|M|=1$ cross-sections could be measured, then a clear choice could 
be made between the predictions of the Born approximation and the exact 
calculation. 

Figures 12 through 16 show our results for the 2-phonon states. In all 
cases, the cross-section predicted by the exact calculation far exceeds 
the predictions of the Born approximation. The natural explanation is 
that these states are predominantly excited by multi-step processes.

\subsubsection{Comparison with the harmonic vibrator model}

Some analyses of multiphonon excitations have used the harmonic 
vibrator model, in which it is assumed that the total effect of the 
interaction on the target protons can be imitated by an operator that 
is linear in the components of the collective variable ${\bf R}_{{\rm 
pn}}$. This has the consequence that multi-phonon states are populated 
by a series of on-shell transitions, and thus the fact that the $\Delta 
M=0$ on-shell transition matrix element becomes small at high 
bombarding energy implies that the $\Delta M=0$ matrix elements will 
play little role in multiphonon excitation. It was shown in Reference 
\cite{BerBau} that the ratios of the cross-sections for the 2-phonon 
states (020), (021), (022), (100) are predicted by this model to be 
1/0/3/2. The absence of population of the state (021)is due to the fact 
that this state consists of phonons with $m=0$ and $m=1$, and the $m=0$ 
phonon is inaccessible. The predictions of the exact calculation, as 
shown in Figures 12 through 16, are quite different. At a bombarding 
energy of 10A GeV, the ratios are approximately 4.1/.3/.1/1.6. The 
important role of the $\Delta M=0$ matrix elements in these numbers is 
evident from the bombarding-energy dependence shown in Figures 12 
through 16. For example, of these four states, only (022) shows a 
cross-section that decreases with bombarding energy, and this is the 
state which could be expected to be least affected by $\Delta M=0$ 
matrix elements. Thus, our microscopic treatment of the interaction, 
which follows the effect of the external pulse on each proton, gives 
quite different results from assumption that the interaction is linear 
in the collective variables. However, it must be borne in mind that the 
error we make by truncating our phonon space at two phonons may 
introduce significant errors into our 2-phonon predictions. This 
requires further investigation.

\subsubsection{The long-wavelength approximation (LWLA)}

It is seen in the cross-section comparisons, Figures 9 through 14,
that at high bombarding energy the LWLA prediction exceeds the result 
of
exact calculation, except for the 1-phonon $|M|=1$ state and the 2-
phonon $|M|=2$ state. The common characteristic of these two
states is that they involve only oscillation in the transverse $({\hat
x},{\hat y})$ directions, whereas all the other final states involve
oscillation in the ${\hat z}$ direction. We have already noted in
connection with Figures 4 and 6 that the excitation of the ${\hat z}$
oscillation takes place rapidly in the vicinity of $t=0$. The LWLA
matrix elements are more strongly peaked in the vicinity of $t=0$ (see
Figures 1b and 2b) due to absence of LWLA dynamic damping. Thus it is
not surprising that the LWLA gives enhanced predictions for states
involving ${\hat z}$ excitation. On the other hand, the excitation of
the ${\hat x},{\hat y}$ oscillation is a more gradual process, covering 
a larger $t$-range, and so is more sensitive to the time region in 
which
the exact matrix elements are greater than the LWLA matrix elements
(Figures 1b, 2b).

\section{Conclusions}

Although we have used a somewhat schematic representation of the 
GDR nuclear states, we can draw some conclusions about the extent to
which the LWLA and Born approximation reproduce the results of our 
exact
calculations. These conclusions may also apply to calculations that use
more realistic nuclear states.

It is apparent from Figures 9 through 16 that at bombarding energies
below about 3A GeV, the LWLA and the exact calculations predict
essentially the same cross-sections, state by state. This observation
supports the validity of several RCE investigations that have used the
LWLA \cite{BCH,LCC}. It is shown in Figure 12 that the agreement
between the LWLA and exact calculations of the {\it total} 1-phonon 
cross-section extends up to about 5A GeV. Above 5A GeV, however, there
are significant deviations between the LWLA and exact cross-sections.
Thus at higher bombarding enrgyies, the LWLA ceases to be a reliable
approximation, especially if it is important to know the cross-sections
for population of individual $M$-states.

Not surprisingly, the Born approximation is unable to predict the 
cross-
section for population of 2-phonon states. Its performance with respect
to the 1-phonon states is analagous to that of the LWLA. At bombarding
energies below 2A GeV, it does well with both $M=0$ and $|M|=1$ states.
At higher energies, it underpredicts the $M=0$ cross-section (Figure
9), and over-predicts the $|M|=1$ cross-section (Figure 10). The net
result is good agreement with the exact prediction for the total
1-phonon cross-section up to about 8A GeV (Figure 11).
\medskip

{\bf Acknowledgements}

The authors wish to thank A. Vitturi for his suggestion that they
investigate the use of numerical Fourier transforms to calculate the
time-dependent interaction matrix elements. One of us (B.F.B)
acknowledges financial support from the INFN.
\medskip
\medskip
\medskip

{\flushleft \Large \bf Appendices}

\appendix
\section*{\large \bf  A: Brink's representation of the GDR phonon 
states}
\label{sec:appa}

\renewcommand{\thesection}{\Alph{section}}
\renewcommand{\theequation}{\thesection.\arabic{equation}}
\setcounter{section}{1}
\setcounter{equation}{0}

We first construct the states of the GDR which are
automatically contained in the
independent particle harmonic 
oscillator shell model. For simplicity, we restrict our attention to 
$N=Z$
closed-shell nuclei. We start by locating the $Z$ protons and the $Z$ 
neutrons
relative to the fixed harmonic oscillator origin by 
${\bf p_{\it 1},\cdots,p_{\it Z},~n_{\it 1},\cdots,n_{\it Z}}$. To 
construct GDR states, we introduce
new position variables
\begin{eqnarray}
{\bf R}&=&\frac{{\bf p_{\it 1}+\cdots+p_{\it Z}~+~n_{\it 
1}+\cdots+n_{\it Z}}}
{2Z}\nonumber\\
{\bf R}_{{\rm pn}}&=&\Bigl(\frac{\bf p_{\it 1}+\cdots+p_{\it 
Z}}{Z}\Bigr)~-~
\Bigl(\frac{\bf n_{\it 1}+\cdots+n_{\it Z}}{Z}\Bigr)
\end{eqnarray}
and relative variables $\mbox{\boldmath$\pi_{\it 1},\cdots,\pi_{{\it Z-
1}},~~
\nu_{\it 1},\cdots,\nu_{{\it Z-1}}$}$. 
The $\mbox{\boldmath$\pi_{\it 1},\cdots,\pi_{{\it Z-1}}$}$
specify the location of the $Z$ protons relative to their mass center. 

\includegraphics[scale=0.35,angle=0]{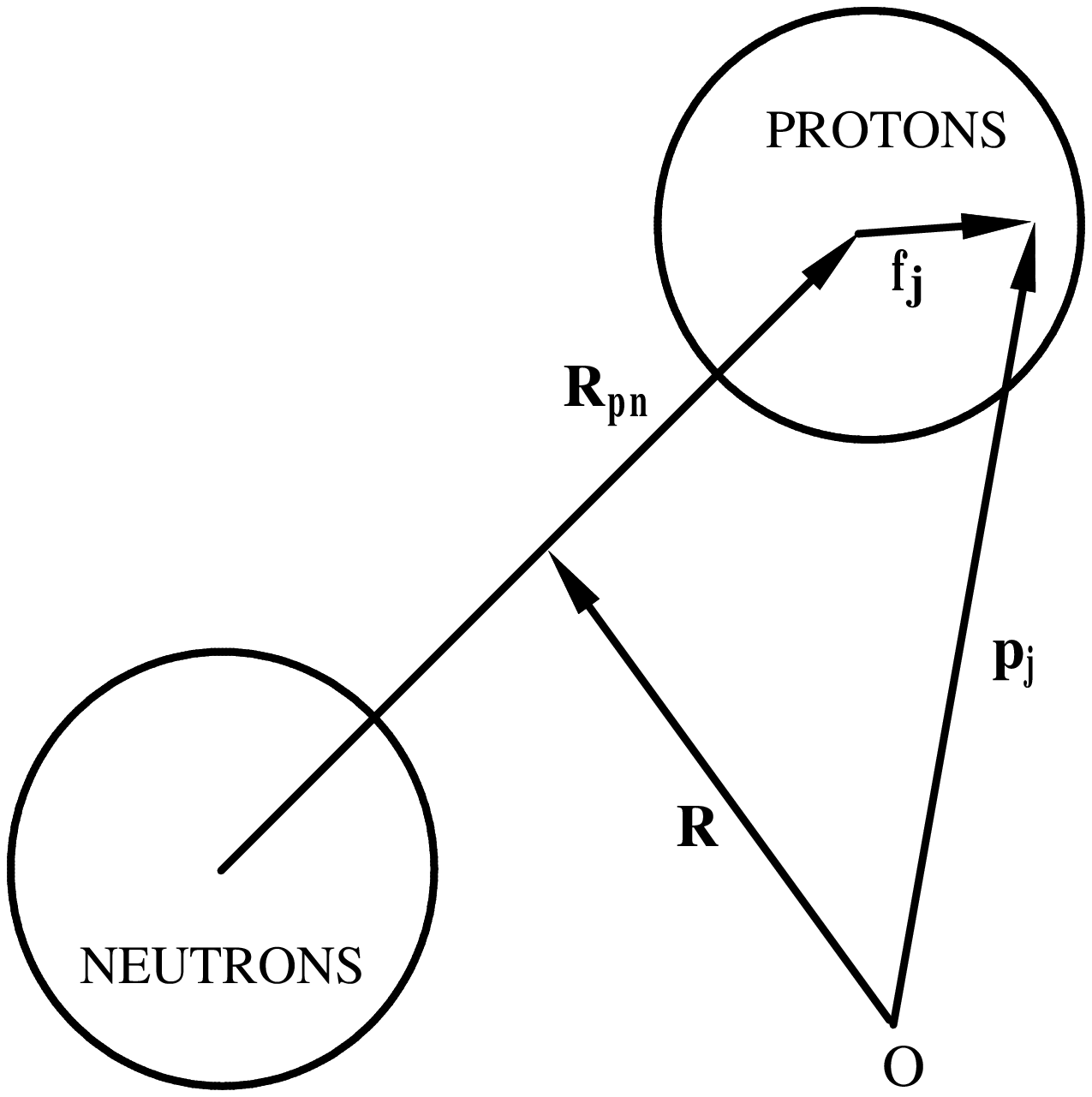}

Figure A1. ${\bf R}_{{\rm pn}}$ locates the position of the proton
mass-center relative to the neutron mass center, whereas ${\bf R}$
locates the total mass center. The $j^{\rm th}$ proton is located
relative to the proton mass center by ${\bf f}_j$.
\vspace{1 cm}

It is 
clear from Fig.(A1) that the $j$th proton is located relative to the
origin by
\beq
{\bf p}_j~=~{\bf R}~+~\frac{1}{2}{\bf R}_{{\rm pn}}~+~{\bf f}_j
(\mbox{\boldmath$\pi_{\it 1},\cdots,\pi_{{\it Z-1}}$})
\eeq
We will not have to specify the
$\mbox{\boldmath$\pi_{\it 1},\cdots,\pi_{{\it Z-1}},~~
\nu_{\it 1},\cdots,\nu_{{\it Z-1}}$}$ or the ${\bf f}_j$ any further; 
we will only
need the ${\bf R}$ and the ${\bf R}_{{\rm pn}}$ dependence of ${\bf 
p}_j$.

If the independent-particle harmonic oscillator Hamiltonian is 
expressed in
terms of these new variables, it separates as follows:
\begin{eqnarray}
H&=&\Bigl(~-\frac{\hbar^2}{2\cdot(2Zm)}~\nabla^2_{\bf R}~+~
\frac{2Zm\omega_{{\rm sm}}^2}{2}~{\rm R}^2~\Bigr)
~+~\Bigl(~-\frac{\hbar^2}{2(\frac{Z}{2}m)}~\nabla^2_{{\bf R}_{{\rm 
pn}}}~+~
\frac{\frac{Z}{2}m\omega_{{\rm sm}}^2}{2}~{\rm R}^2_{{\rm pn}}~\Bigr)
\nonumber\\
&+&H'(\mbox{\boldmath$\pi_{\it 1},\cdots,\pi_{{\it Z-1}},~~
\nu_{\it 1},\cdots,\nu_{{\it Z-1}}$})~~~.
\end{eqnarray}
Here $m$ is the nucleon mass and $\omega_{{\rm sm}}$ is the single-
nucleon 
oscillator
frequency. Corresponding to this separability, we can write our 
harmonic 
oscillator eigenstates in the form
\beq
\Psi_\alpha~=~\Psi^{N_1,L_1}_{M_1}({\bf R})~\Psi^{N_2,L_2}_{M_2}({\bf 
R}_
{{\rm pn}})
~\times~\chi(\mbox{\boldmath$\pi_{\it 1},\cdots,\pi_{{\it Z-1}},~~
\nu_{\it 1},\cdots,\nu_{\rm Z-1}$})~~~.
\eeq
The spin degrees of freedom are incorporated in $\chi$, in such a way 
as to produce a state which is antisymmetric with respect to the 
exchange 
of any two protons or any two neutrons. Note that changing 
$N_1,~L_1,~M_1$
or $N_2,~L_2,~M_2$ has no effect on the antisymmetry, since the 
variables
${\bf R}$ and ${\bf R}_{{\rm pn}}$ are symmetric with respect to the 
exchange 
of any two protons or any two neutrons.

It is clear that the state of lowest energy, consistent with the Pauli
principle, will have zero quanta in the variables ${\bf R}$ and ${\bf 
R}_
{{\rm pn}}$:
\beq
\Psi_{\rm ground-state}~=~\Psi^{0,0}_0({\bf R})~\Psi^{0,0}_0({\bf R}_
{{\rm pn}})~
~\times~\chi_0(\mbox{\boldmath$\pi_{\it 1},\cdots,\pi_{\it Z-1},~~
\nu_{\it 1},\cdots,\nu_{\it Z-1}$})~~.
\eeq
Now let us consider a series of states of the form
\beq
\Psi_M^{N,L}~=~\Psi^{0,0}_0({\bf R})~\Psi^{N,L}_M({\bf R}_{{\rm pn}})~
~\times~\chi_0(\mbox{\boldmath$\pi_{\it 1},\cdots,\pi_{\it Z-1},~~
\nu_1,\cdots,\nu_{Z-1}$})~~.
\eeq
These states all have no quanta of center-of-mass motion. The relative
motions of the protons to each other, and of the neutrons to each 
other, 
are exactly the same as they are in the shell-model ground state.
The only feature distinguishing the states (A6) is the motion of the 
proton
mass-center relative to the neutron mass-center. Brink \cite{Bri} 
identified 
these states as belonging to the GDR, because this collective 
oscillation
of the protons relative to the neutrons was the original interpretation
given by Goldhaber and Teller \cite{GTel} to the phenomenon now known 
as the 
GDR.

It can be seen from (A3) that the energy associated with the ${\bf R}_
{{\rm pn}}$
quanta is $\hbar\omega_{{\rm sm}}$, the same as the energy associated 
with the 
individual oscillator quanta. It was suggested by Wilkinson \cite{Wil}, 
and 
proved by Brown and Bolsterli \cite{Bro}, that the residual interaction 
between the nucleons will increase the GDR phonon energy.

The harmonic oscillator radial functions associated with the individual
nucleons were specified in (4.6). The same radial functions are 
used for the
GDR states $\Psi^{N,L}_M({\bf R}_{{\rm pn}})$, except that $m$ 
is replaced by $Zm/2$, which is the reduced mass for
the relative motion of the proton and neutron spheres.

Electromagnetic properties of states are normally discussed in the 
long-wavelength approximation, in which the electric dipole operator 
has the form
$$
{\bf {\cal M}}=\sqrt{\frac{3}{4 \pi}}~\sum_{j=1}^Z {\bf p}_{{\rm j}}
$$
Equation (A.2) can be used to express this operator in terms of ${\bf 
R, R_{{\rm pn}}}, {\mbox{\boldmath$\pi$}}_{j}$. The only part of this 
operator that has matrix elements between states of the GDR band (A.6) 
is
$$
{\bf {\cal M}}=\sqrt{\frac{3}{4 \pi}}~\frac{Z}{2}~{\bf R}_{{\rm pn}}
$$
which can only connect states which differ by one GDR phonon. Thus the 
E1 transition between the ground state and the 1-phonon Brink level 
exhausts the E1 sum rule.

\section*{\large \bf B: Contribution of the scalar potential}
\label{sec:appb}
\setcounter{section}{2}
\setcounter{equation}{0}

It is shown in \cite{BZ} that if the nuclear 
states $\phi_\alpha,~\phi_\beta$ are defined with ``time-reversal'' 
phases,
\beq
(\phi^{I_\alpha}_{M_\alpha})^*~=~(-1)^{I_\alpha-M_\alpha}
~\phi^{I_\alpha}_{M_{-\alpha}}~~,
\eeq
then the $V_{\beta \alpha}(\omega)$ will be purely real. We will use
this phase convention.

The charge density $\rho_{\beta \alpha}({\bf r'})$ is defined in terms 
of the 
nuclear eigenstates $\phi_\beta,\phi_\alpha$ by
\begin{equation}
\rho_{\beta \alpha}({\bf r'})=e~\int~d^3p_1...d^3n_Z
\sum_{j=1}^Z~\delta({\bf r'-p_{\it j}})
\phi^*_\beta\phi_\alpha
\end{equation}
The sum extends over the protons only.
By using (B2) and (4.3) in (4.1), we see that the scalar potential 
contribution
to $V_{\beta\alpha}(\omega)$ is 
\beq
\int d^3r'~\varphi_{_C}^{\rm ret}({\bf r'},\omega)\rho_{\beta 
\alpha}({\bf r'})
~=~\sum_{\lambda
\mu}C_{\lambda,\mu}(\omega)K_\mu(\frac{|\omega|b}{v \gamma})
~\int d^{3}r'~\rho_{\beta \alpha}({\bf
r'})j_{\lambda}(\frac{|\omega|}{c}r')Y_{\mu}^{\lambda}({\hat r'})~~.
\eeq
The integral required here also occurs in the Fourier transform of
$\rho_{\beta\alpha}({\bf r'})$ :
\beq
\rho_{\beta\alpha}({\bf k})~\equiv~\int~e^{i{\bf k\cdot r'}}~
\rho_{\beta \alpha}({\bf r'})d^3r'
~=~4\pi~\sum_{\lambda\mu}~i^\lambda~Y^{\lambda *}_\mu({\hat{\bf k}})
~\int d^{3}r'~\rho_{\beta \alpha}({\bf r'})   
j_{\lambda}(kr')Y_{\mu}^{\lambda}
({\hat r'}) 
\eeq
if we set $k=|\omega|/c$. Thus we turn our attention to the evaluation 
of
$\rho_{\beta\alpha}({\bf k})$:
$$
\rho_{\beta \alpha}({\bf k})~\equiv~\int~e^{i{\bf k\cdot r'}}~
\rho_{\beta \alpha}({\bf r'})d^3r'
~=~\int d^3r'e^{i{\bf k\cdot r'}}
<\phi_\beta|e\sum_{j=1}^Z~\delta({\bf r'-p_{\it j}})|\phi_\alpha>
$$
\beq
=~<\phi_\beta|e\sum_{j=1}^Z~e^{i{\bf k\cdot p'}_j}|\phi_\alpha>~~~.
\eeq
We now choose states in the GDR band (A6) for the $\phi_\beta$ and 
$\phi_\alpha$, and use (A2) to express ${\bf p}_j$ in terms of the new 
variables:
\begin{eqnarray*}
\rho_{\beta\alpha}({\bf k})&=&e\int~d^3R|\Psi^{0,0}_0|^2~\times~
\int d^3R_{{\rm pn}}\Psi^{*N_\beta,L_\beta}_{M_\beta}({\bf R}_{{\rm 
pn}})
\Psi^{N_\alpha,L_\alpha}_{M_\alpha}({\bf R}_{{\rm pn}})\\
&\times&\int~d^3\mbox{\boldmath$\pi_{\it 1},\cdots,{\it d^3}\nu_{\it Z-
1}$}
|\chi_0(\mbox{\boldmath$\pi_{\it 1},\cdots,\pi_{\it Z-1},~~
\nu_{\it 1},\cdots,\nu_{\it Z-1}$})|^2\\
&\times&\sum_{j=1}^Z~e^{i{\bf k\cdot(R+\frac{1}{2} R_{\it pn}
+f_{\it j}(\mbox{\boldmath$\pi_{\it 1},\cdots,\pi_{\it Z-1},~~
\nu_{\it 1},\cdots,\nu_{\it Z-1}$})}}
\end{eqnarray*}
\beq
~=~Q\int d^3R_{{\rm pn}}\Psi^{*N_\beta,L_\beta}_{M_\beta}({\bf R}_{{\rm 
pn}})
~e^{i\frac{{\bf k}}{2}\cdot{\bf R}_{{\rm pn}}}
\Psi^{N_\alpha,L_\alpha}_{M_\alpha}({\bf R}_{{\rm pn}})
\eeq
with
\begin{eqnarray}
Q&\equiv&e\int~d^3R|\Psi^{0,0}_0|^2~e^{i{\bf k\cdot R}}
~\times~\int~d^3\mbox{\boldmath$\pi_{\it 1},\cdots,{\it d^3}\nu_{\it Z-
1}$}
|\chi_0(\mbox{\boldmath$\pi_{\it 1},\cdots,\pi_{\it Z-1},~~
\nu_{\it 1},\cdots,\nu_{\it Z-1}$})|^2\nonumber\\
&\times&\sum_{j=1}^Z~e^{i{\bf k\cdot
f_{\it j}(\mbox{\boldmath$\pi_{\it 1},\cdots,\pi_{\it Z-1},~~
\nu_{\it 1},\cdots,\nu_{\it Z-1}$})}}
\end{eqnarray}
Note that the factor $Q$ is independent of the GDR quantum numbers. 
This
is a consequence of our assumption that the proton and neutron spheres
remain unexcited during the ${\bf R}_{{\rm pn}}$ oscillation. Thus we 
can 
calculate $Q$ by considering the ground-state proton density
\beq
\rho_{00}({\bf k})~=~e\int~d^3r'~e^{i{\bf k\cdot r'}}~\frac{1}{4\pi}
\sum_{n',\ell}~2(2\ell+1)u^2_{n',\ell}(r')
~=~e\sum_{n',\ell}~2(2\ell+1)~\int_0^\infty~u^2_{n',\ell}(r')j_0(kr')r'
2dr'
\eeq
The $n',\ell$ sum extends over the single-particle levels occupied in 
the
ground state (for example: 00, 01, 02, 10, for $^{40}$Ca). Using (B6) 
to
evaluate $\rho_{00}({\bf k})$ yields
\beq
\rho_{00}({\bf k})~=~e^{-\frac{k^2}{8Z\nu}}~\times~Q~~~,
\eeq
so that
\beq
Q=e^{\frac{k^2}{8Z\nu}}\times e~\sum_{n,\ell}~2(2\ell+1)~\int_0^\infty~
u^2_{n,\ell}(r')j_0(kr')r'^2dr'~~.
\eeq

To evaluate the ${\bf R}_{{\rm pn}}$ integral in (B6), we write
$$
\Psi^{*N_\beta,L_\beta}_{M_\beta}({\bf R}_{{\rm pn}})
\Psi^{N_\alpha,L_\alpha}_{M_\alpha}({\bf R}_{{\rm pn}})
~=~(-1)^{L_\beta-M_\beta}~\Psi^{N_\beta,L_\beta}_{-M_\beta}({\bf R}_
{{\rm pn}})
\Psi^{N_\alpha,L_\alpha}_{M_\alpha}({\bf R}_{{\rm pn}})
$$
\beq
=(-1)^{L_\beta-M_\beta}~\sum_L~(L_\beta L_\alpha~-M_\beta M_\alpha|
L~M_\alpha-M_\beta) \times
\Bigl[~\Psi^{N_\beta,L_\beta}({\bf R}_{{\rm pn}})
\Psi^{N_\alpha,L_\alpha}({\bf R}_{{\rm pn}})~\Bigr]^L_{M_\alpha-
M_\beta}.
\eeq
We then use the Talmi-Moshinsky coefficients, defined by
\begin{eqnarray}
\Bigl[~\psi^{n_1\ell_1}({\bf r}_1)\psi^{n_2\ell_2}({\bf 
r}_2)~\Bigr]^L_M
&=&\sum_{n,\lambda,N,\Lambda}~i^{\ell_1+\ell_2-\lambda-\Lambda}
~(n_1\ell_1,
n_2\ell_2|n\lambda,N\Lambda)_L\nonumber\\
&&\Bigl[~\psi^{n\lambda}(\frac{{\bf r}_{1}-{\bf
r}_{2}}{\sqrt{2}}) ~\psi^{N\Lambda}(\frac{{\bf r}_{1}+{\bf
r}_{2}}{\sqrt{2}})~\Bigr]^L_M
\end{eqnarray}
with ${\bf r}_1={\bf r}_2={\bf R}_{{\rm pn}}$, to write
$$
\Bigl[\Psi^{N_\beta,L_\beta}({\bf R}_{{\rm pn}})
\Psi^{N_\alpha,L_\alpha}({\bf R}_{{\rm pn}})\Bigr]^L_{M_\alpha-M_\beta}
~~~~~~~~~~~~~~~~~~~~~~~~~~~~~~~~~~~~~~~~~~~~~~~~~~~~~~~~~~~~~~~~~
$$
$$
=~\sum_{n,\lambda,N,\Lambda}i^{L_\alpha+L_\beta-\lambda-\Lambda}
~(N_\beta L_\beta,N_\alpha L_\alpha|n\lambda,N\Lambda)_L
~\Bigl[~\Psi^{n\lambda}
(0) ~\psi^{N\Lambda}(\sqrt{2}{\bf R}_{{\rm pn}})~\Bigr]^{L}
_{M_\alpha-M_\beta}~~~~~~~~~~~~~
$$
\beq
=~i^{\ell_1+\ell_2-L}(\frac{Z\nu}{2\pi})^{3/4}~\sum_{n,N}~(N_\beta 
L_\beta,
N_\alpha L_\alpha|n0,NL)_L
\times \sqrt{\frac{(2n+1)!!}{2n!!}}~
\Psi^{NL}_{M_\alpha-M_\beta}(\sqrt{2}{\bf R}_{{\rm pn}})
\eeq
The extra factor $i^{\ell_1+\ell_2-\lambda-\Lambda}$ is due to our use
of time-reversal phases. The $n,N$ sum in (B13) is restricted by the 
quanta-preserving condition 
$2(n+N)+L=2(N_\beta+N_\alpha)+L_\beta+L_\alpha$,
which in turn requires that $L_\beta+L_\alpha-L$ be even.

To Fourier transform $\Psi^{NL}_{M_\alpha-M_\beta}(\sqrt{2}{\bf R}_
{{\rm pn}})$
we use the general harmonic oscillator result
$$
\frac{1}{(2\pi)^{3/2}}~\int_{-\infty}^\infty~ e^{i{\bf k\cdot r}}
\psi^{n \ell}_m({\bf r})d^3r~=~i^{2n+\ell}~{\tilde \psi}^{n\ell}_m({\bf 
k})~,
$$
where ${\tilde \psi}^{n\ell}_m({\bf k})$ has a size parameter ${\tilde 
\nu}$
which is the reciprocal of the size parameter of $\psi^{n 
\ell}_\mu({\bf r})$.
Then
$$
\int_{-\infty}^\infty~ e^{i\frac{\bf k}{2}\cdot{\bf R}_{{\rm pn}}}
\Psi^{NL}_{M_\alpha-M_\beta}(\sqrt{2}{\bf R}_{{\rm pn}})d^3{\bf R}_
{{\rm pn}}
~=~i^{2N+L}~\pi^{3/2}{\tilde \Psi}^{NL}_{M_\alpha-M_\beta}(\frac{{\bf 
k}}
{2\sqrt{2}})
$$
\beq
=~(-1)^{N+L}~\pi^{3/2}~Y^L_{M_\alpha-M_\beta}({\hat{\bf k}})
{\tilde u}_{NL}(\frac{k}{2\sqrt{2}}).
\eeq
By combining (B6), (B8), (B13) and (B14), we get 
\begin{eqnarray}
\rho_{\beta\alpha}({\bf k})&=&(-1)^{M_\beta}i^{L_\alpha-L_\beta}
(\frac{Z\nu\pi}{2})^{3/4}
~e^{\frac{k^2}{8Z\nu}}~\sum_{n',\ell}2(2\ell+1)
 \int_0^\infty~j_0(kr')
u^2_{n',\ell}(r')r'^2dr'\nonumber\\
&&\times~\sum_{n,N,L}i^{2N+L}(L_\beta L_\alpha~-M_\beta M_\alpha|
L~M_\alpha-M_\beta)\nonumber\\
&&\times~(N_\beta L_\beta,
N_\alpha L_\alpha|n0,NL)_L
~\times\sqrt{\frac{(2n+1)!!}{2n!!}}
~{\tilde u}_{NL}(\frac{k}{2\sqrt{2}})Y^L_{M_\alpha-M_\beta}({\hat{\bf 
k}})
\end{eqnarray}

Finally, comparison with (B3) and (B4) shows that
\begin{eqnarray}
&&\int d^3r'~\varphi_{_C}^{\rm ret}({\bf r'},\omega)\rho_{\beta \alpha}
({\bf r'})
=\frac{(-1)^{M_\alpha}i^{L_\alpha-L_\beta}}{4\pi}
~(\frac{Z\nu\pi}{2})^{3/4}
~e^{\frac{(|\omega/c)^2}{8Z\nu}}K_\mu(\frac({|\omega|b}{v\gamma})\nonumber\\
&&\times~\sum_{n,\ell}2(2\ell+1)~\int_0^\infty~j_0(\frac{|\omega|}{c}r'
)~
u^2_{n,\ell}(r')r'^2dr'
~\sum_L~C_{L,M_\beta-M_\alpha}~(L_\beta L_\alpha~-M_\beta M_\alpha|
L~M_\alpha-M_\beta)\nonumber\\
&&\times~\sum_{n,N}(-1)^N~(N_\beta L_\beta,
N_\alpha L_\alpha|n0,NL)_L
~\sqrt{\frac{(2n+1)!!}{2n!!}}
~{\tilde u}_{NL}(\frac{\frac{|\omega|}{c}}{2\sqrt{2}}),  
\end{eqnarray}
which gives us an explicit expression for the contribution of the 
scalar 
potential to $V_{\beta \alpha}(\omega)$. This expression is summarized 
in (4.5)

\section*{\large \bf C: Contribution of the vector potential}
\label{sec:appc}
\setcounter{section}{3}
\setcounter{equation}{0}

The analysis is similar to the one presented in Appendix B for the 
scalar
potential. We need 
\begin{eqnarray}
-\frac{1}{c}~\int d^3r'~{\bf j}_{\beta \alpha}({\bf r'}) \cdot 
{\bf A}_{_C}^{\rm ret}({\bf r'},\omega)
&=&-\frac{v}{c^2}~\int~d^3r'
[{\bf j}_{\beta \alpha}({\bf r'})]_z\varphi^{\rm ret}_{_C}
({\bf r'},\omega)\nonumber\\
&=&-\frac{v}{c^2}\sum_{\lambda,\mu}~C_{\lambda,\mu}~\int d^{3}r~
j_{\lambda}(kr')Y_{\mu}^{\lambda}({\hat{\bf r'}})
[{\bf j}_{\beta \alpha}({\bf r'})]_z ~.
\end{eqnarray}
To obtain these integrals we study the Fourier transform of 
$[{\bf j}_{\beta \alpha}({\bf r'})]_z$:
\begin{eqnarray}
[{\bf j}_{\beta \alpha}({\bf k})]_z&\equiv&\int~e^{i{\bf k\cdot r'}}~
[{\bf j}_{\beta \alpha}({\bf r'})]_z~d^3r'
~=~4\pi\sum_{\lambda,\mu}~i^\lambda Y_{\mu}^{*\lambda}({\hat{\bf k}})
~\int d^{3}r~j_{\lambda}(kr')
[{\bf j}_{\beta \alpha}({\bf r'})]_z\nonumber\\
&=&\int d^{3}r~e^{i{\bf k\cdot r'}}\int~d^3p_1\cdots d^3n_Z
~\frac{e\hbar}{2mi}\sum_{j=1}^Z~\delta({\bf
r'-p}_j)
~\Bigl(\phi^*_\beta \frac{\partial}{\partial
p_{j,z}}\phi_\alpha~-~\phi_\alpha \frac{\partial}{\partial
p_{j,z}}\phi^*_\beta\Bigr)\nonumber\\ 
&=&\int~d^3p_1,\cdots
d^3n_Z~\sum_{j=1}^Z~e^{i{\bf k\cdot p_j}}
\times \Bigl(\phi^*_\beta\frac
{\partial}{\partial p_{j,z}}\phi_\alpha~-~\phi_\alpha
\frac{\partial}{\partial p_{j,z}}\phi^*_\beta\Bigr)
\end{eqnarray}
We now take $\phi_\beta,~\phi_\alpha$ to be states in the GDR band 
(A6),
and transform to the variables ${\bf R, R}_{{\rm pn}},
\mbox{\boldmath$\pi_{\it 1},\cdots,\pi_{{\it Z-1}},~~\nu_{\it 
1},\cdots,
\nu_{{\it Z-1}}$}$. To do
this we use (A2) to obtain
$$
\frac{\partial}{\partial p_{j,z}}~=~\frac{1}{2Z}\frac{\partial}
{\partial {\rm R}_z}
~+~\frac{1}{Z}\frac{\partial}{\partial {\rm R}_{{\rm pn},z}}~+
~\sum_{k=1\to j-1;\mu=x,y,z}
\Bigl(~\frac{\partial}{\partial p_{j,z}}\pi_{k,\mu}~\Bigr)
~\frac{\partial}{\partial \pi_{k,\mu}}
$$
Since the $\mbox{\boldmath$\pi_{\it k}$}$ involve only the relative 
positions
of the ${\bf p}_j$, the quantities 
$\frac{\partial}{\partial p_{j,z}}\pi_{k,\mu}$ are independent of ${\bf 
R}$ 
and ${\bf R}_{{\rm pn}}$

Equation (C2) now takes the form 
$$
[{\bf j}_{\beta \alpha}({\bf 
k})]_z~=~\frac{e\hbar}{2(Zmi)}Q\times\int~d^3
R_{{\rm pn}}
~e^{\frac{{\bf k\cdot R}_{{\rm pn}}}{2}}
$$
\beq
\times
\Bigl(\Psi^{*N_\beta L_\beta}_{M_\beta}
({\bf R}_{{\rm pn}})\frac{\partial}{\partial R_{{\rm pn},z}}
\Psi^{N_\alpha L_\alpha}_{M_\alpha}
({\bf R}_{{\rm pn}})~-~\Psi_{M_\alpha}^{N_\alpha L_\alpha}
\frac{\partial}{\partial R_{{\rm pn},z}}\Psi^{*N_\beta 
L_\beta}_{M_\beta}
({\bf R}_{{\rm pn}})~\Bigr)
\eeq
where $Q$
is the quantity defined in (B7) and evaluated in (B10). Note that (C3) 
vanishes
for diagonal matrix elements ($N_\beta=N_\alpha,~L_\beta=L_\alpha,~
M_\beta=M_\alpha$). The derivatives in (C3) can be expressed as linear
combinations of harmonic oscillator wave functions using the relation
$$
\frac{\partial}{\partial Z}\Psi^{n,\ell}_m({\bf 
R})~=~\sqrt{\frac{\nu}{2}}
~\times
$$
$$
\Bigl[~\sqrt{\frac{(\ell-m)(\ell+m)}{(2\ell-1)(2\ell+1)}}
\bigl[~\sqrt{2n+2\ell+1} \Psi^{n,\ell-1}_m({\bf R})~+~\sqrt{2n+2}
\Psi^{n+1, \ell-1}_m({\bf R})~\bigr]
$$
$$
-~\sqrt{\frac{(\ell+1-
m)(\ell+1+m)}{(2\ell+1)(2\ell+3)}}\bigl[~\sqrt{2n}
\Psi^{n-1,\ell+1}_m({\bf R})+\sqrt{2n+2\ell+3}\Psi^{n,\ell+1}_m({\bf 
R})~\bigr]
~\Bigl]
$$
This converts (C3) into a linear combination of terms such as (B6), 
which
can be evaluated exactly as was done in Appendix B.

\section*{\large \bf D: Symmetries}
\label{sec:appd}
\setcounter{section}{4}
\setcounter{equation}{0}

Since the projectile moves in the $y-z$ plane, its scalar and vector 
potentials
will be invariant under reflection across the $y-z$ plane. Therefore,
only reflection-symmetric target states will be excited from the 
reflection-symmetric ground state. These are
$$
\Psi^{NL}_0,~~~~~~~~~~~~~{\rm or}~~~~~~~~~~ 
\frac{1}{\sqrt{2}}(\Psi^{NL}_{|M|}+\Psi^{NL}_{-|M|}).
$$
Thus if $M_\alpha\neq 0$ and $M_\beta\neq 0$, we must calculate
$$
<\frac{1}{\sqrt{2}}(\Psi^{N_\beta L_\beta}_{M_\beta}+
\Psi^{N_\beta L_\beta}_{-M_\beta})|V(\omega)|
\frac{1}{\sqrt{2}}(\Psi^{N_\alpha L_\alpha}_{M_\alpha}+
\Psi^{N_\alpha L_\alpha}_{-M_\alpha})>
$$
$$
=~<\Psi^{N_\beta L_\beta}_{|M_\beta|}|V(\omega)|
\Psi^{N_\alpha L_\alpha}_{|M_\alpha|}>+
<\Psi^{N_\beta L_\beta}_{|M_\beta|}|V(\omega)|
\Psi^{N_\alpha L_\alpha}_{-|M_\alpha|}>~~.
$$
If $M_\beta\neq 0$ and $M_\alpha=0$, we must calculate 
$$
<\frac{1}{\sqrt{2}}(\Psi^{N_\beta L_\beta}_{M_\beta}+
\Psi^{N_\beta L_\beta}_{-M_\beta})|V(\omega)|
\Psi^{N_\alpha L_\alpha}_0>~=~
\sqrt{2}<\Psi^{N_\beta L_\beta}_{|M_\beta|}|V(\omega)|
\Psi^{N_\alpha L_\alpha}_0>~~.
$$
It can be verified that Equation B3 of Reference \cite{BZ}
$$
<\Psi^{N_\beta L_\beta}_{M_\beta}|V(\omega)|
\Psi^{N_\alpha L_\alpha}_{M_\alpha}>~=~
<\Psi^{N_\alpha L_\alpha}_{M_\alpha}|V(-\omega)|
\Psi^{N_\beta L_\beta}_{M_\beta}>,
$$
is satisfied by the matrix elements discussed in Appendices B and C.

\medskip
\medskip
\medskip

{\large \bf Figure Captions}
\medskip

Figure 1. (a) $V_{000-011}(\omega)$ for a bombarding energy of 10A GeV 
and
an impact parameter of 12 fm. The peak at $\omega=0$ is narrower for 
the
exact curve than for the long-wavelength approximation because the 
exact
curve shows the effect of dynamic damping. 
(b) $V_{000-011}(t)$ corresponding to the $V_{000-011}(\omega)$
of Figure 1a. The narrower $\omega$ dependence of the exact curve
results in a broader $t$ dependence.
\vspace{.5cm}

Figure 2. (a) The same as Figure 1a, but for an impact parameter of 50 
fm.
(b) The same as Figure 1b, but for an impact parameter of 50 fm.
\vspace{.5cm}

Figure 3. Excitation probability of the state $\frac{1}{\sqrt{2}}
[\Psi^{01}_{1}+\Psi^{01}_{-1}]$ as a function of
$t$, for a bombarding energy of 10A GeV and an impact parameter of 200
fm. The Born approximation excitation probability is shown for
comparison.
\vspace{.5cm}

Figure 4. The same as Figure 3, except that the state is $\Psi^{01}_0$. 
The $t
\rightarrow \infty$ prediction is $0.6 \times 10^{-5}$, compared to a
Born approximation prediction of about $10^{-8}$.
\vspace{.5cm}

Figure 5. The same as Figure 3, but for an impact parameter of 12 fm.
\vspace{.5cm}

Figure 6. The same as Figure 4, but for an impact parameter of 12 fm.
\vspace{.5cm}

Figure 7. The asymptotic excitation probability of the state 
$\frac{1}{\sqrt{2}}
[\Psi^{01}_{1}+\Psi^{01}_{-1}]$ as a
function of impact parameter, for a bombarding energy of 10A GeV.
\vspace{.5cm}

Figure 8. The same as Figure 7, except that the state is 
$\frac{1}{\sqrt{2}}
[\Psi^{02}_{2}+\Psi^{02}_{-2}]$ . Because
of the importance of multistep processes in the excitation of this
state, the Born approximation underestimates the excitation probability
even at large impact parameter.
\vspace{.5cm}

Figure 9.  Cross-section for excitation of the state $\Psi^{01}_0$ as a 
function of the bombarding energy.
\vspace{.5cm}

Figure 10. Cross-section for excitation of the state 
$\frac{1}{\sqrt{2}}
[\Psi^{01}_{1}+\Psi^{01}_{-1}]$ as a function of the bombarding energy.
\vspace{.5cm}

Figure 11. Total cross-section for excitation of the one-phonon states
(the sums of the curves in Figures 9 and 10).
\vspace{.5cm}

Figure 12.  Cross-section for excitation of the state $\Psi^{02}_0$ as 
a 
function of the bombarding energy. The maximum Born cross-section is 
0.038 mb.
\vspace{.5cm}

Figure 13. Cross-section for excitation of the state 
$\frac{1}{\sqrt{2}}
[\Psi^{02}_{1}+\Psi^{02}_{-1}]$ as a function of the bombarding energy.
\vspace{.5cm}

Figure 14. Cross-section for excitation of the state 
$\frac{1}{\sqrt{2}}
[\Psi^{02}_{2}+\Psi^{02}_{-2}]$ as a function of the bombarding energy.
\vspace{.5cm}

Figure 15.  Cross-section for excitation of the state $\Psi^{10}_0$ as 
a 
function of the bombarding energy. The maximum Born cross-section is 
0.0006 mb.
\vspace{.5cm}

Figure 16. Total cross-section for excitation of the two-phonon states
(the sums of the curves in Figures 12, 13, 14, and 15). The maximum 
Born 
cross-section is 0.65 mb.
\vspace{.5cm}

\pagebreak
\includegraphics[scale=.30,angle=0]{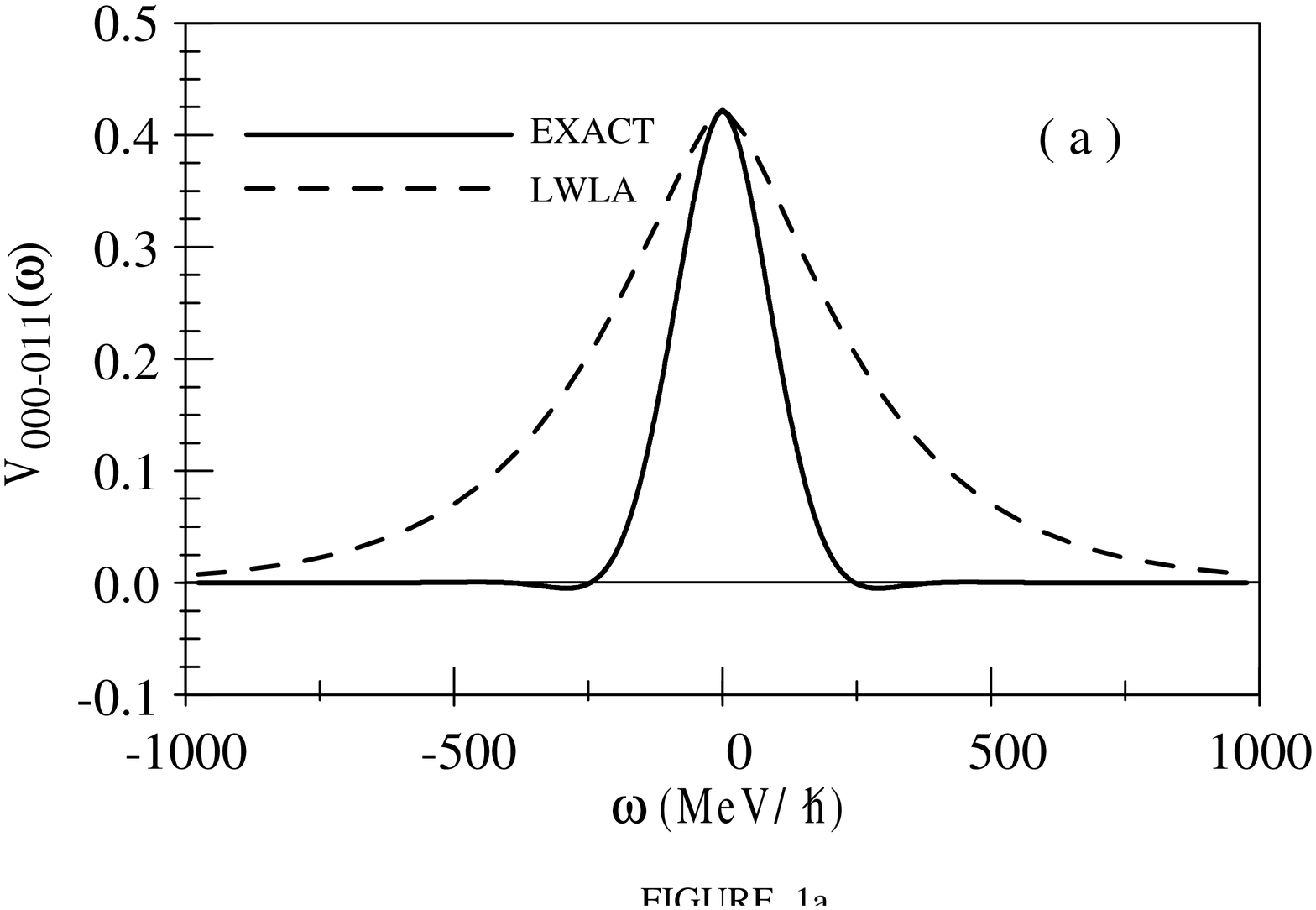}
\includegraphics[scale=.30,angle=0]{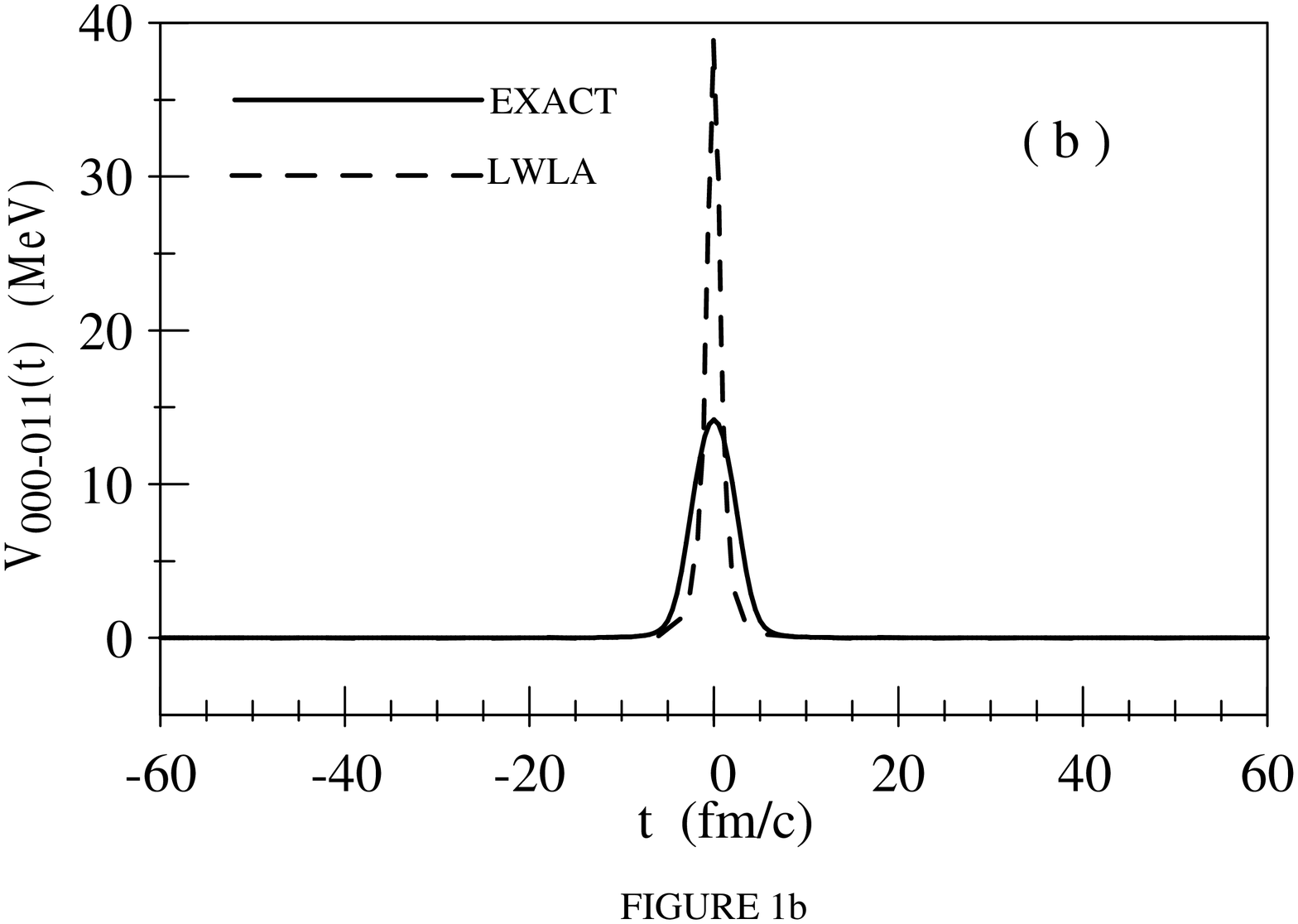}
\includegraphics[scale=.30,angle=0]{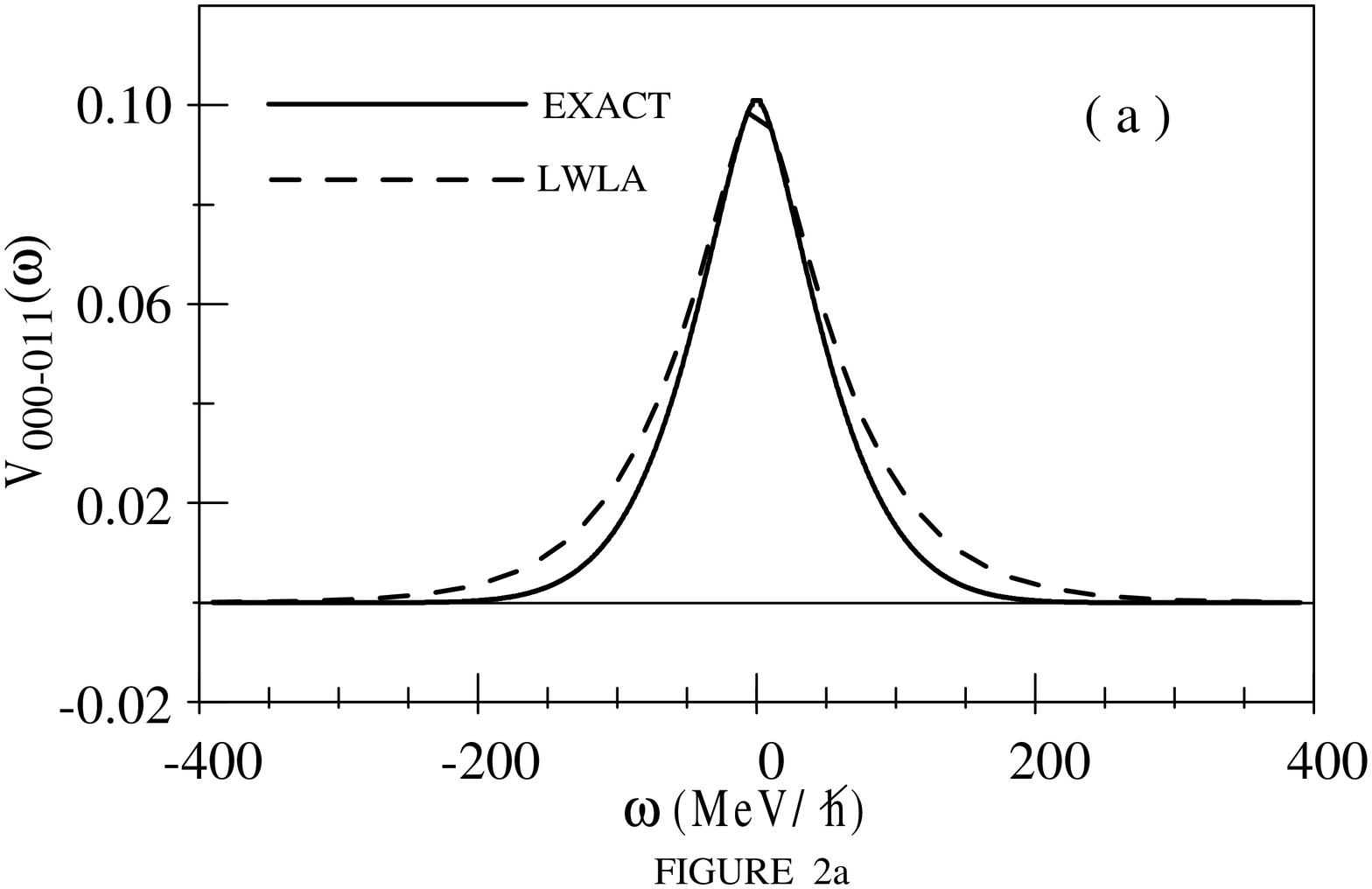}
\includegraphics[scale=.30,angle=0]{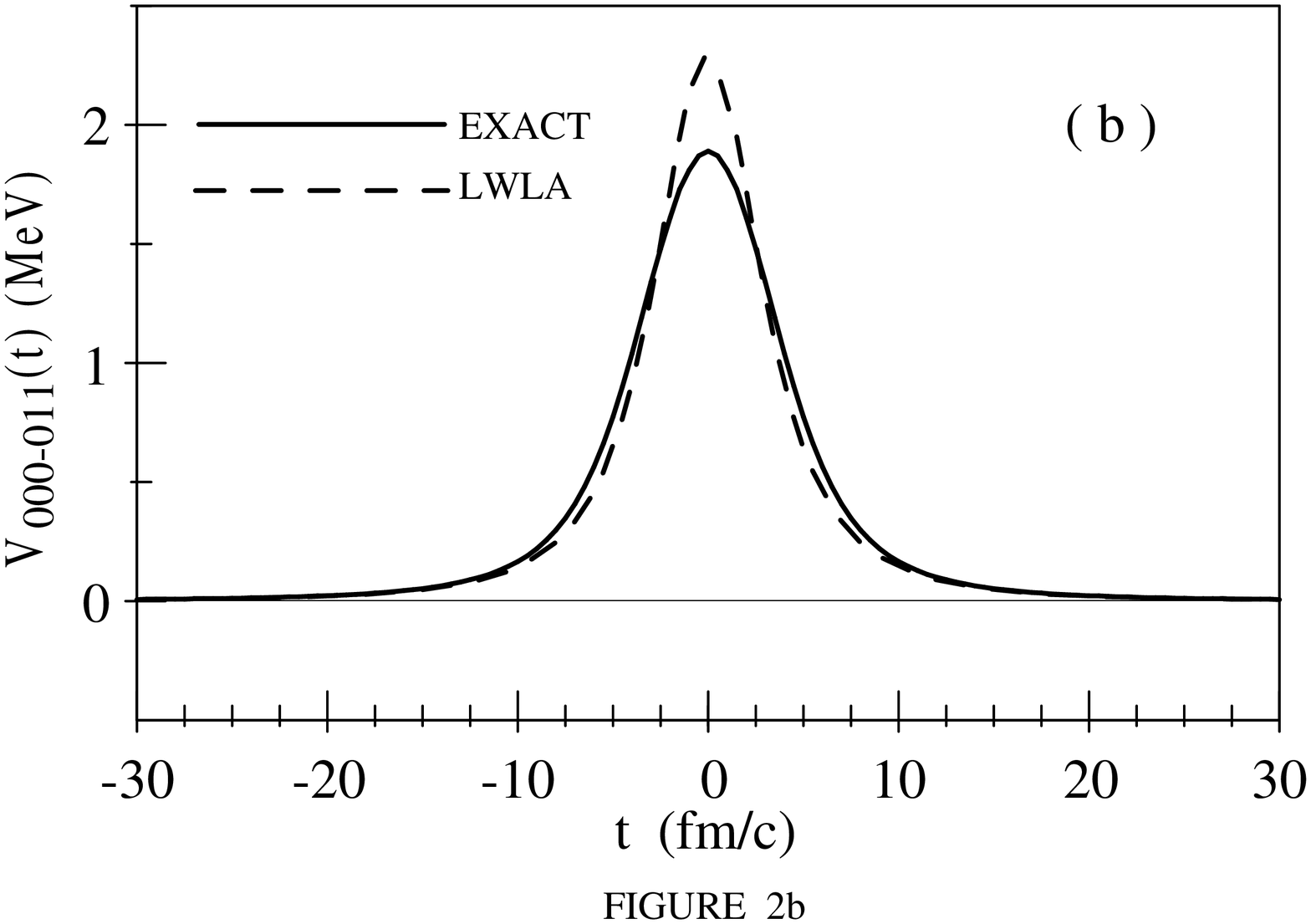}
\includegraphics[scale=.30,angle=0]{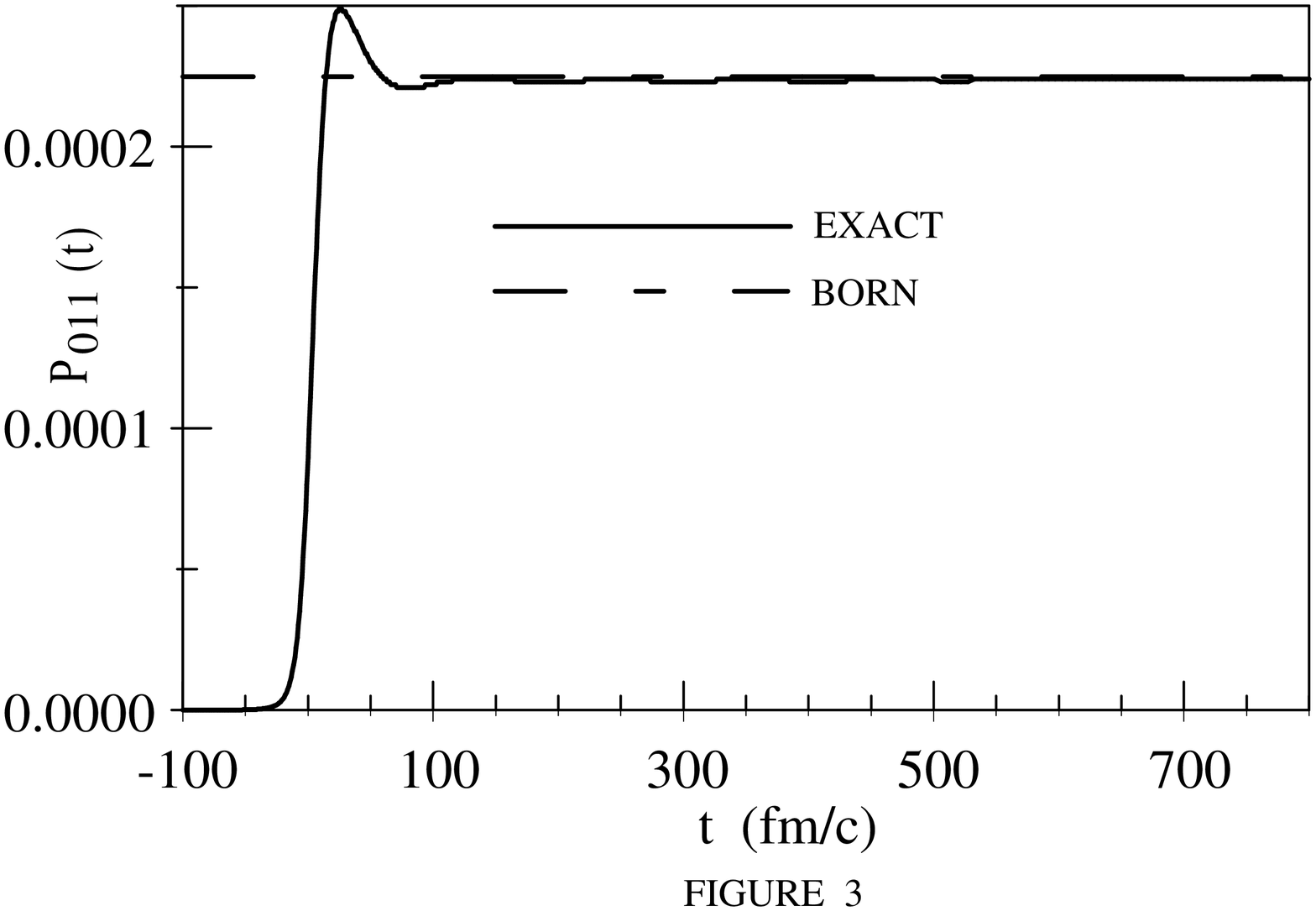}
\includegraphics[scale=.30,angle=0]{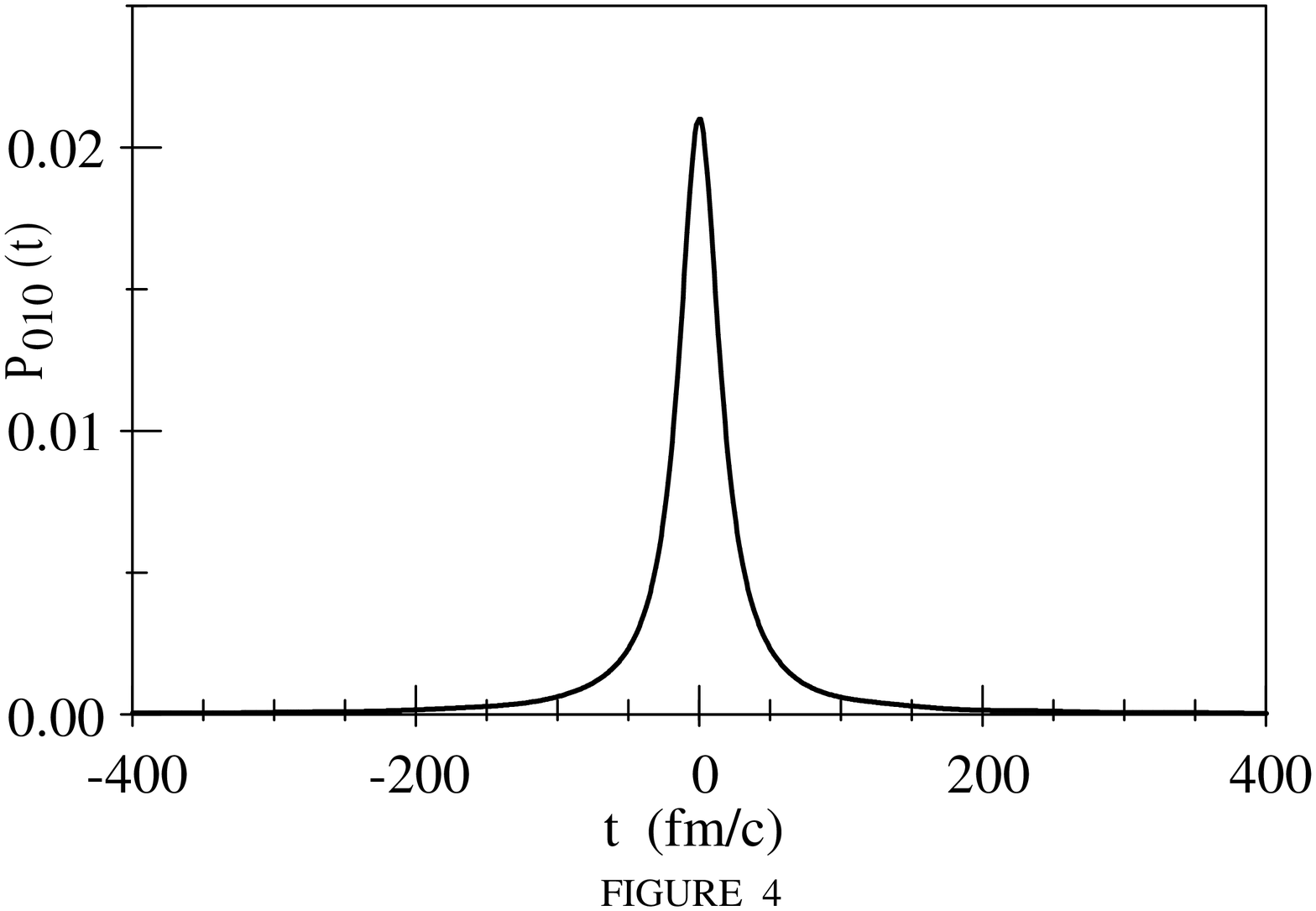}
\includegraphics[scale=.30,angle=0]{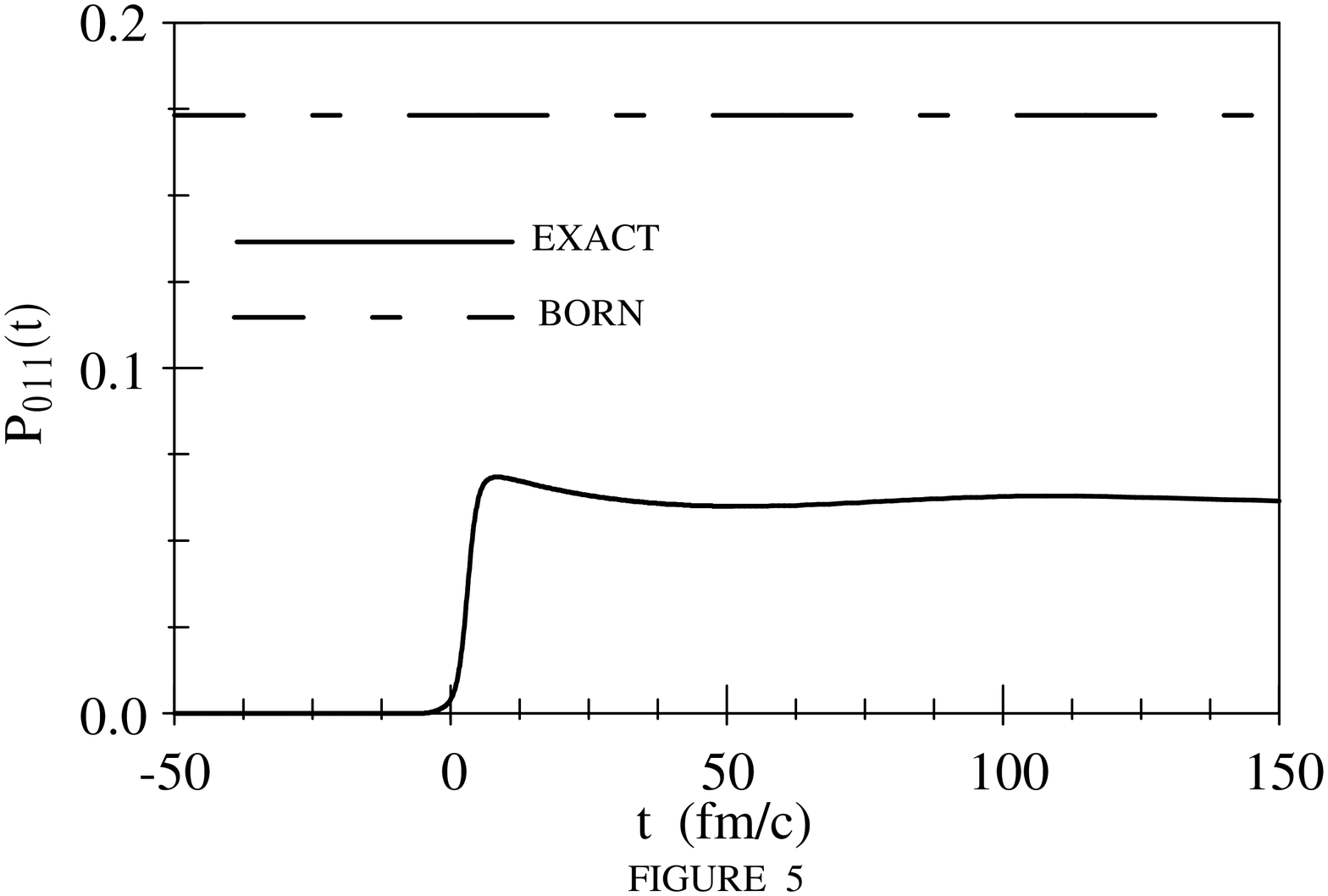}
\includegraphics[scale=.30,angle=0]{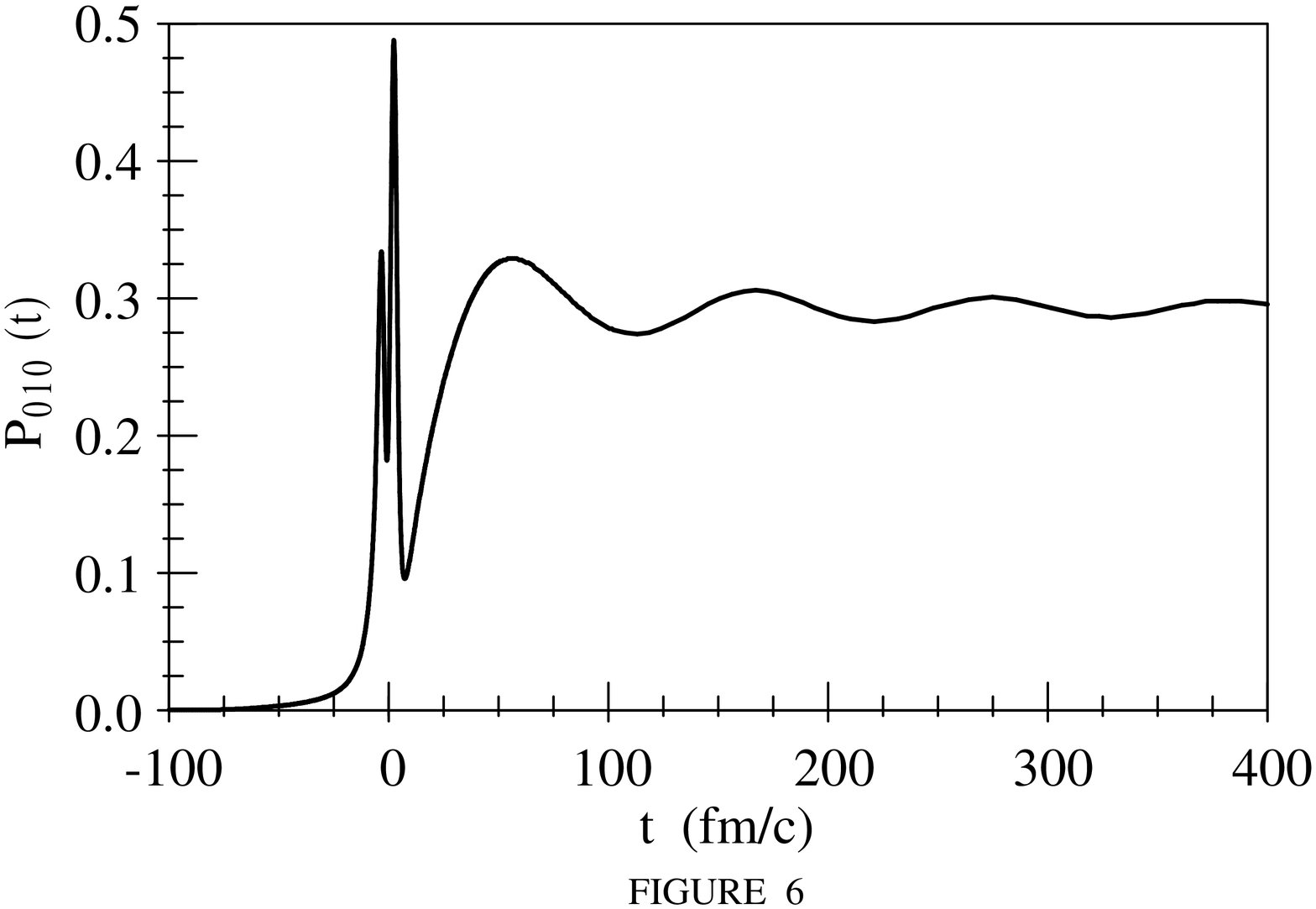}
\includegraphics[scale=.30,angle=0]{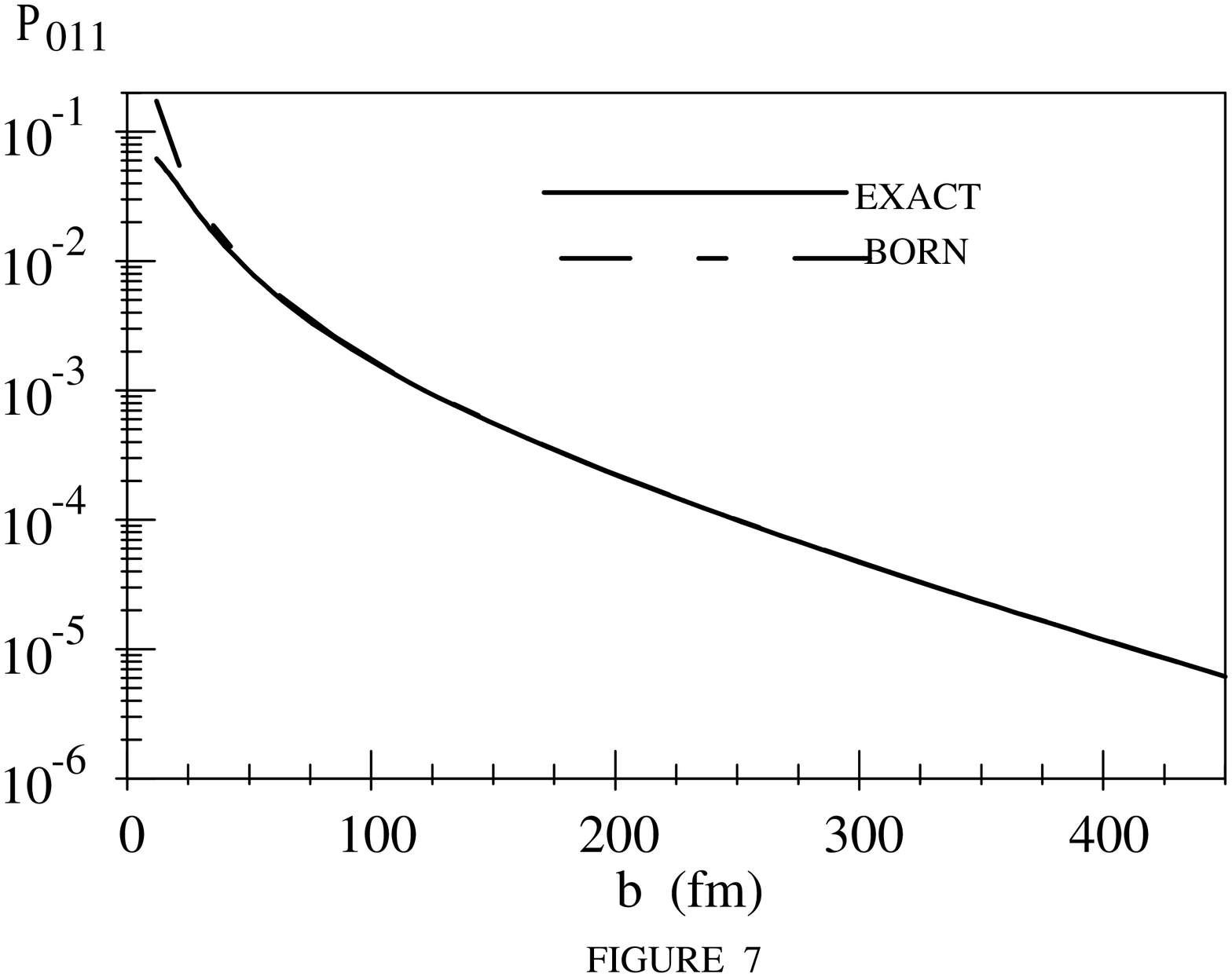}
\includegraphics[scale=.30,angle=0]{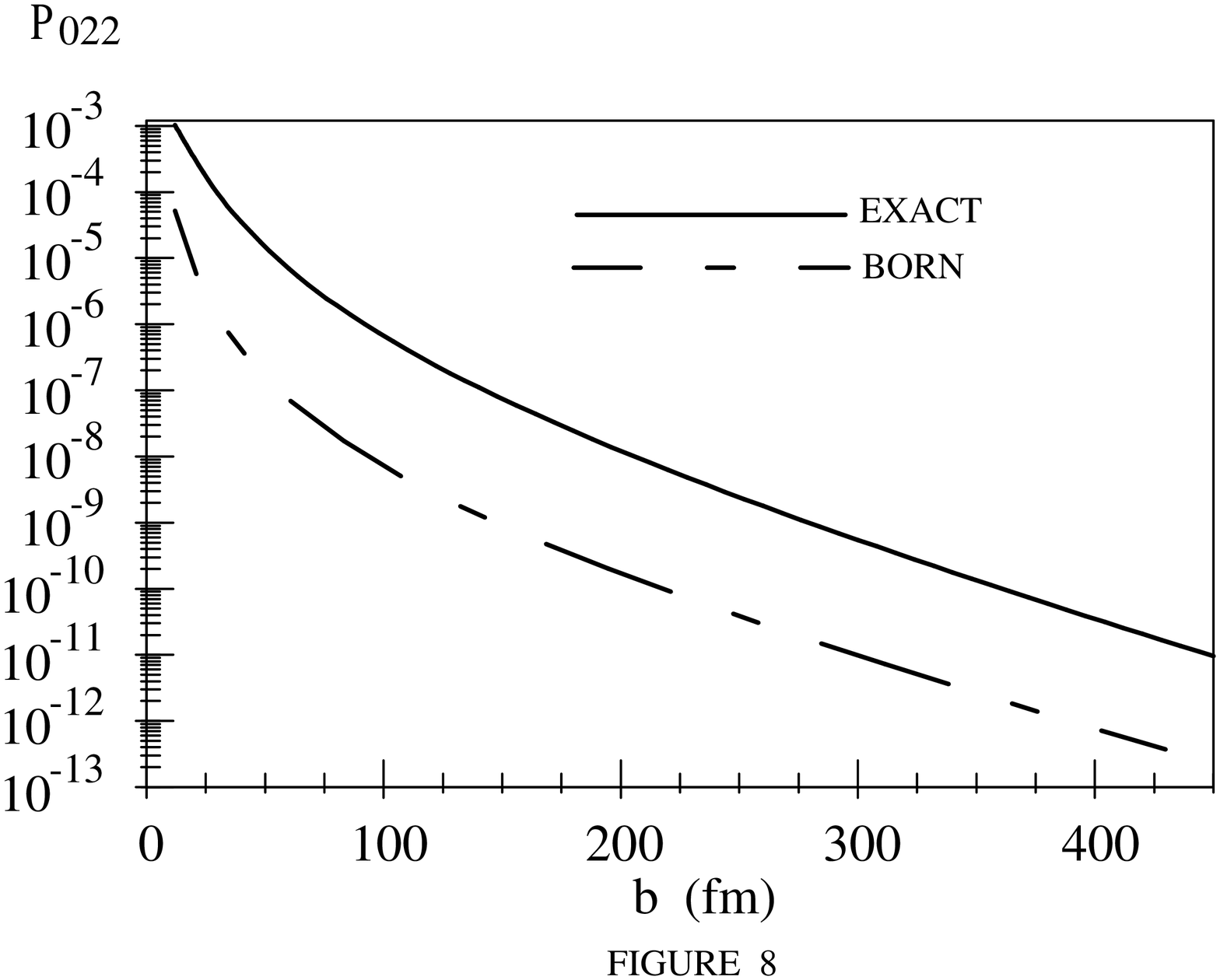}
\includegraphics[scale=.30,angle=0]{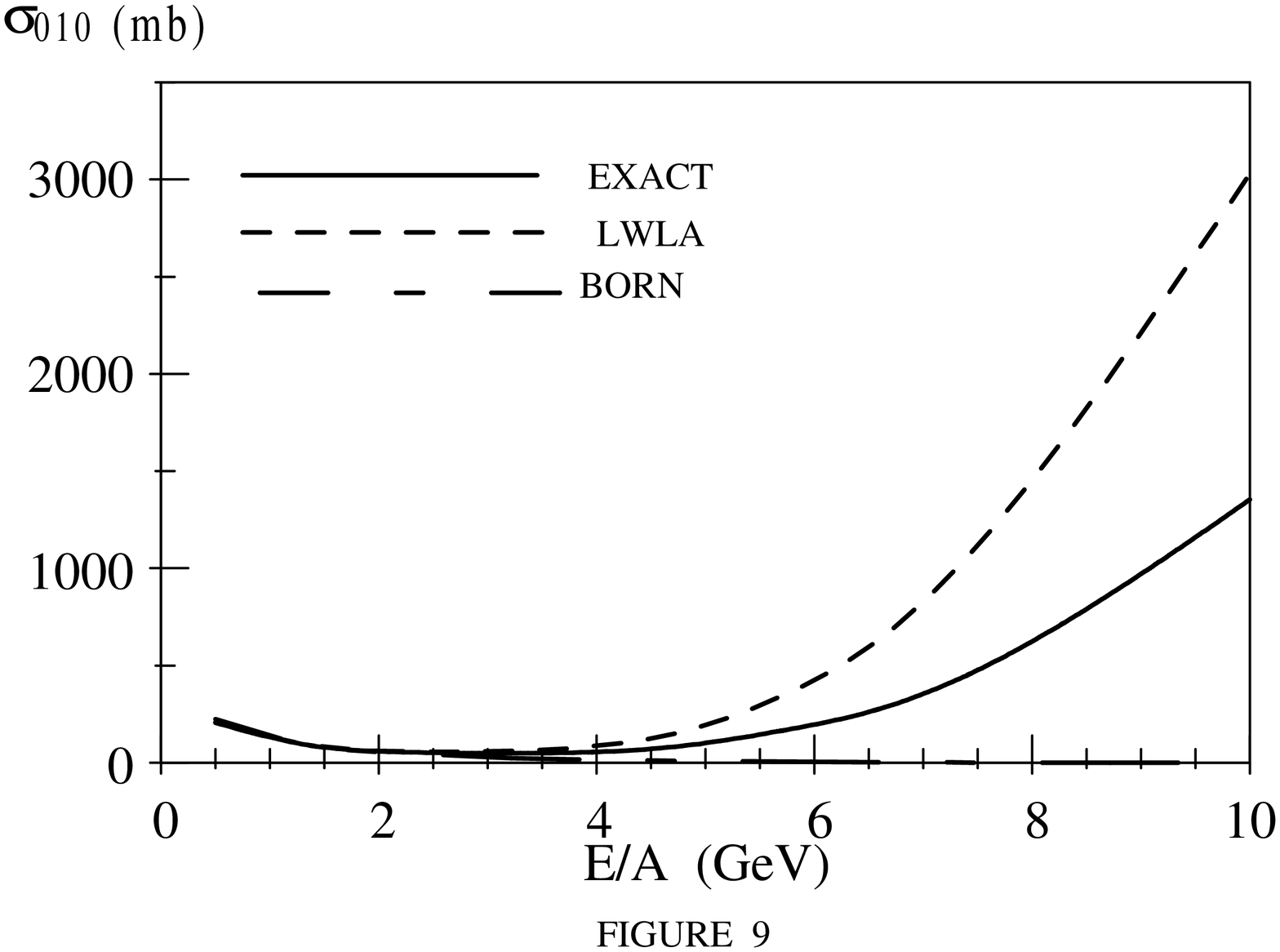}
\includegraphics[scale=.30,angle=0]{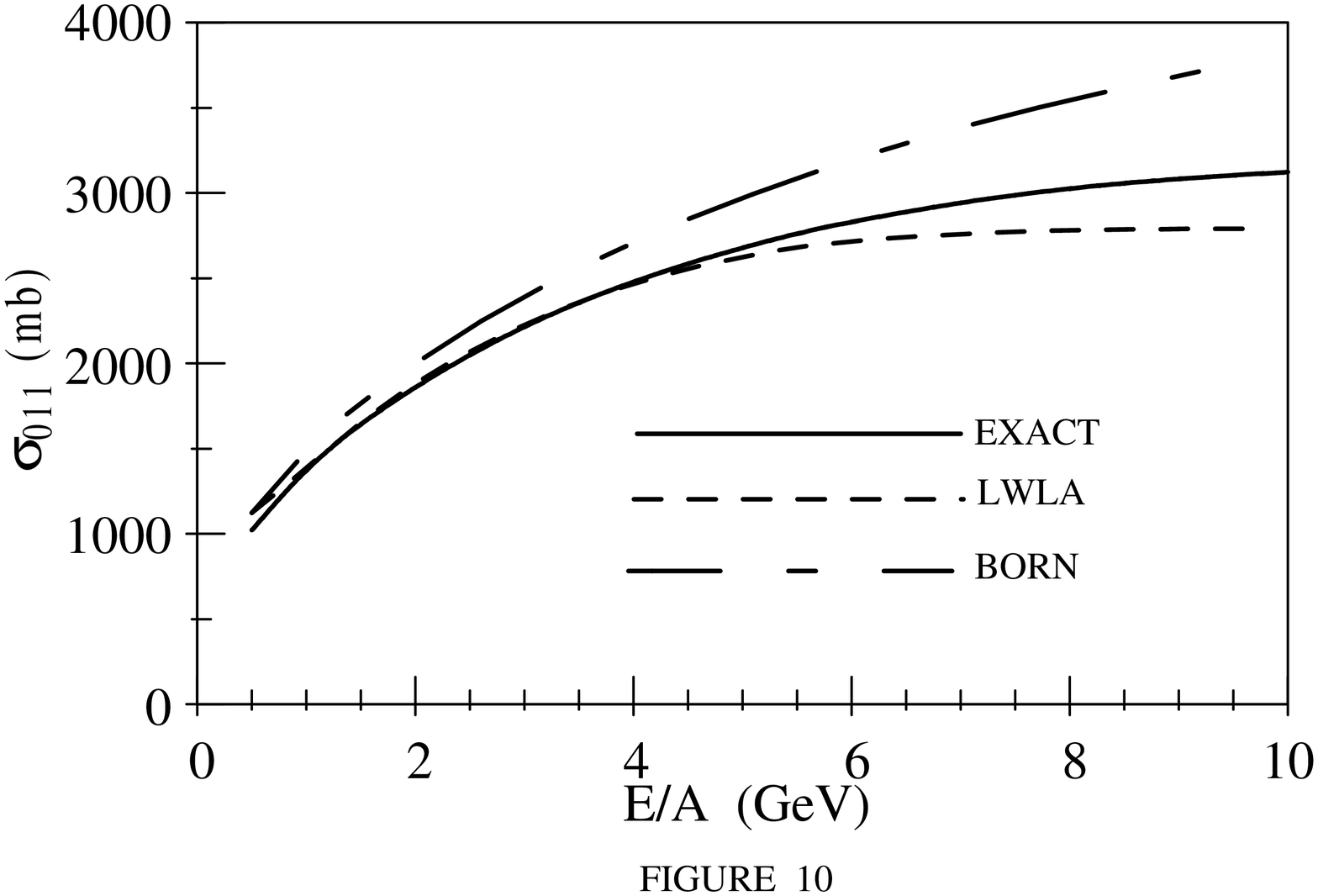}
\includegraphics[scale=.30,angle=0]{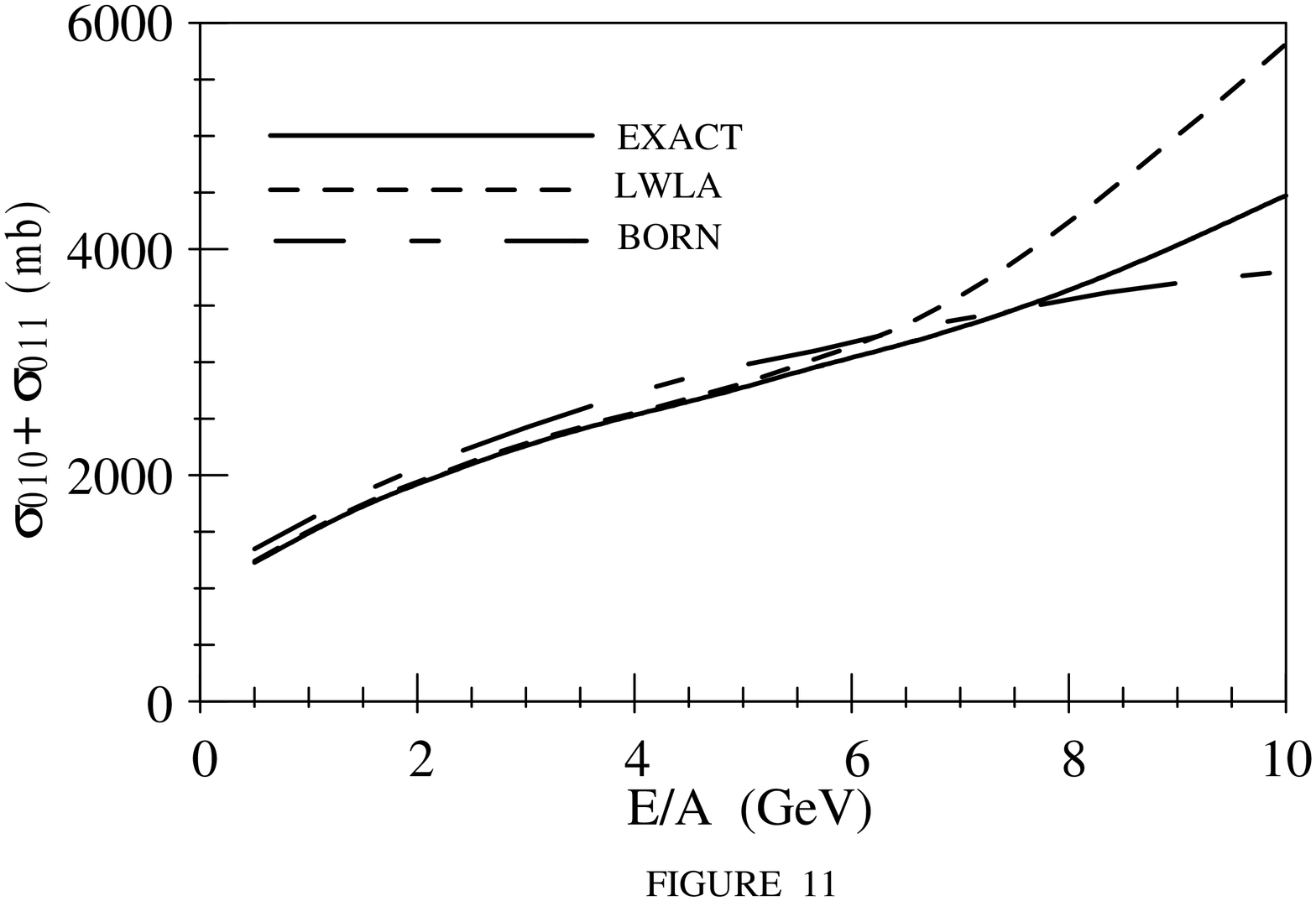}
\includegraphics[scale=.30,angle=0]{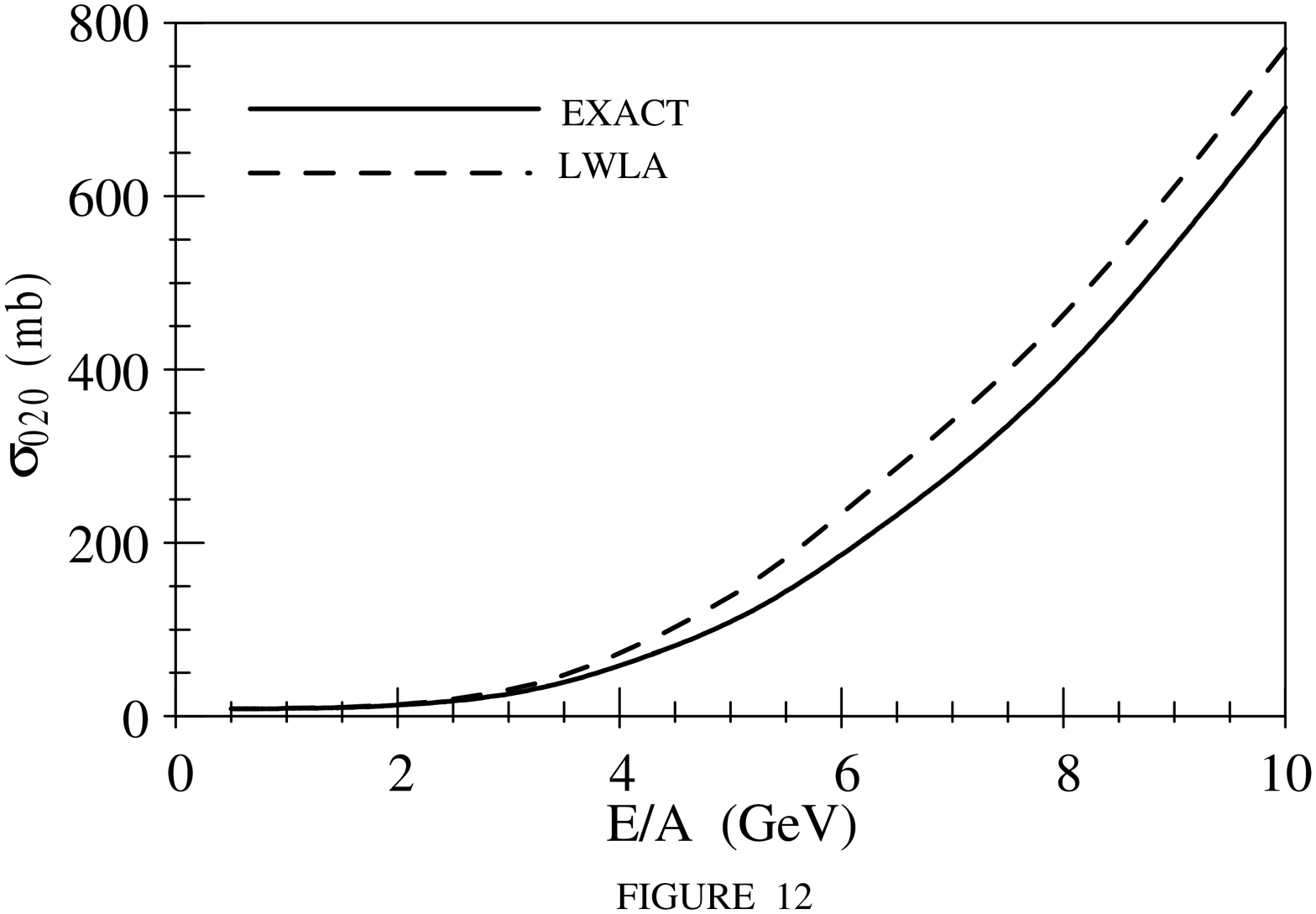}
\includegraphics[scale=.30,angle=0]{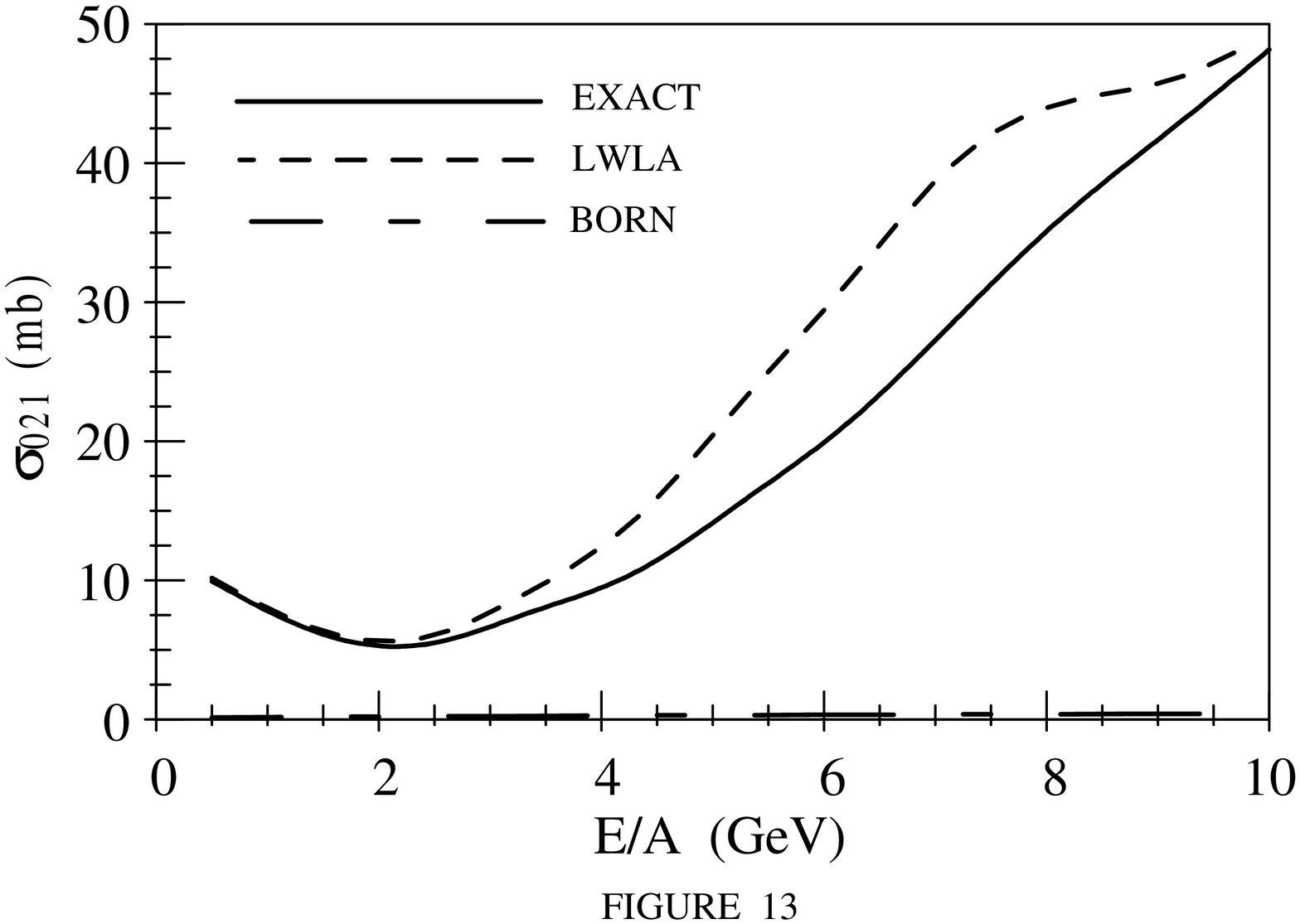}
\includegraphics[scale=.30,angle=0]{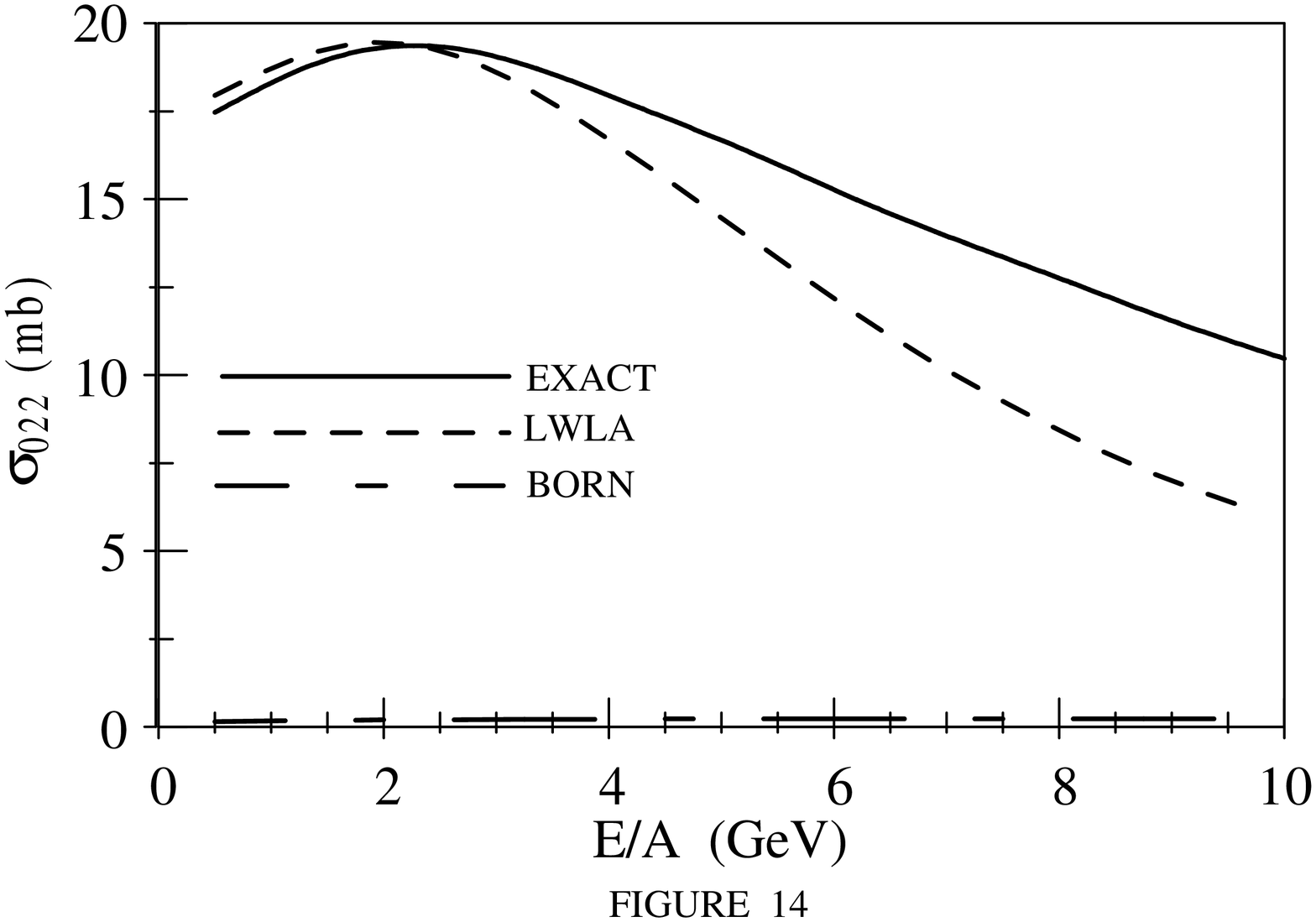}
\includegraphics[scale=.30,angle=0]{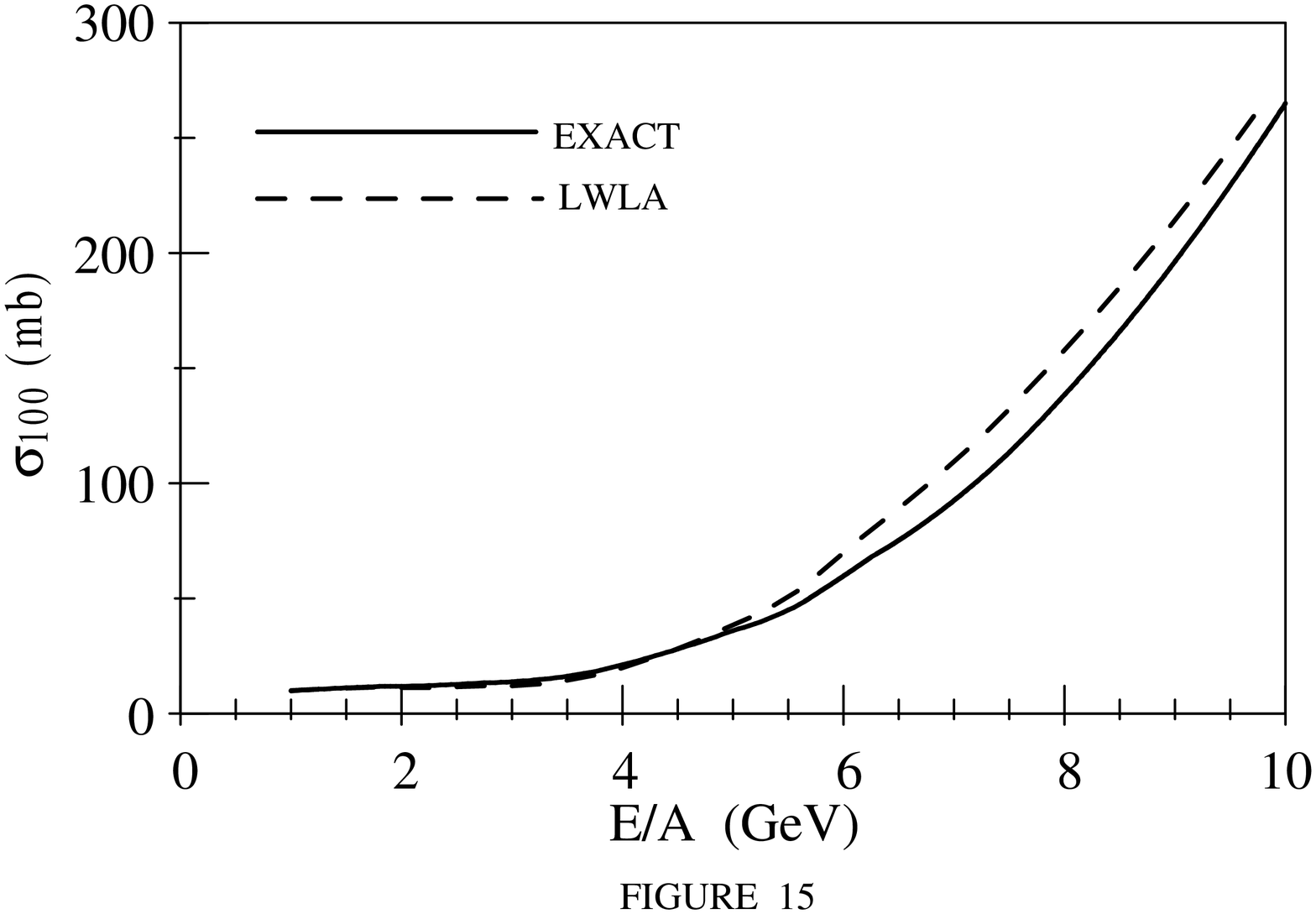}
\includegraphics[scale=.30,angle=0]{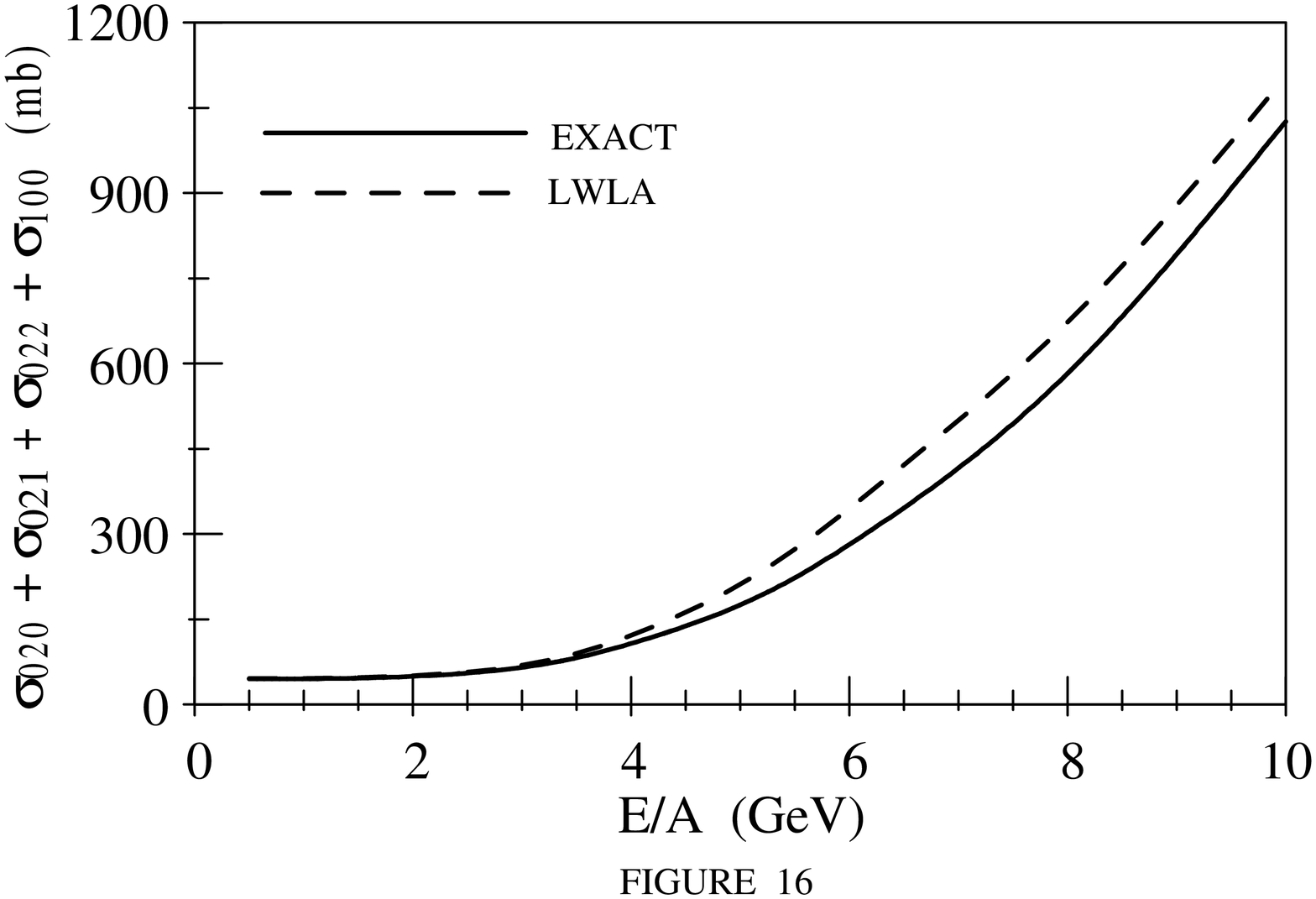}

\end{document}